\pdfoutput=1
\documentclass[fleqn,usenatbib,useAMS]{mnras}

\usepackage{amssymb}	
\usepackage{multicol}        
\usepackage{bm}		
\usepackage{pdflscape}	
\usepackage[T1]{fontenc}
\usepackage{ae,aecompl}

\usepackage {graphicx}
\usepackage{natbib}
\usepackage{aas_journals}
\usepackage{color}
\usepackage{amsmath}
\definecolor{mygray}{gray}{0.7}

\newcommand{\bes}{\begin{equation*}}
\newcommand{\ees}{\end{equation*}}
\newcommand{\bea}{\begin{eqnarray}}
\newcommand{\eea}{\end{eqnarray}}
\newcommand{\beas}{\begin{eqnarray*}}
\newcommand{\eeas}{\end{eqnarray*}}

\newcommand{\mpc}{h^{-1}\mathrm{Mpc}}

\newcommand{\ltsima}{$\; \buildrel < \over \sim \;$}
\newcommand{\lsim}{\lower.5ex\hbox{\ltsima}}
\newcommand{\gtsima}{$\; \buildrel > \over \sim \;$}
\newcommand{\gsim}{\lower.5ex\hbox{\gtsima}}

\def\gtrsim{\mathrel{\hbox{\rlap{\hbox{\lower4pt\hbox{$\sim$}}}\hbox{$>$}}}}
\let\ga=\gtrsim
\def\lesssim{\mathrel{\hbox{\rlap{\hbox{\lower4pt\hbox{$\sim$}}}\hbox{$<$}}}}
\let\la=\lesssim

\definecolor{mygray}{gray}{0.5}

\newcommand{\be}{\begin{equation}}
\newcommand{\ee}{\end{equation}}
\newcommand{\ba}{\begin{eqnarray}}
\newcommand{\ea}{\end{eqnarray}}

\title[ISW with supervoids selected from eBOSS QSO data]{Evidence for a high-$z$ ISW signal from supervoids in the distribution of eBOSS quasars}

\author[Andr\'as Kov\'acs et al.]{A. Kov\'acs$^{1,2}$\thanks{Juan de la Cierva Fellow, Email: akovacs@iac.es}, R. Beck$^{3}$, A. Smith$^4$, G. R\'acz$^3$, I. Csabai$^3$, I. Szapudi$^5$\\
$^{1}$ Instituto de Astrof\'{\i}sica de Canarias (IAC), Calle V\'{\i}a L\'{a}ctea, E-38200, La Laguna, Tenerife, Spain\\
$^{2}$ Departamento de Astrof\'{\i}sica, Universidad de La Laguna (ULL), E-38206, La Laguna, Tenerife, Spain\\
$^{3}$ Department of Physics of Complex Systems, ELTE E\"{o}tv\"{o}s Lor\'and University, Pf. 32, H-1518 Budapest, Hungary\\
$^{4}$ IRFU, CEA, Universit\'e Paris-Saclay, F-91191 Gif-sur-Yvette, France \\
$^{5}$ Institute for Astronomy, University of Hawaii, 2680 Woodlawn Drive, Honolulu, HI, 96822}

\begin{document}
\date{Submitted 2022}
\pagerange{\pageref{firstpage}--\pageref{lastpage}} \pubyear{2022}
\maketitle
\label{firstpage}
\begin{abstract}
The late-time integrated Sachs-Wolfe (ISW) imprint of $R\gtrsim100~\mpc$ super-structures is sourced by evolving large-scale potentials due to a dominant dark energy component in the $\Lambda$CDM model. The aspect that makes the ISW effect distinctly interesting is the repeated observation of stronger-than-expected imprints from supervoids at $z\lesssim0.9$. Here we analyze the un-probed key redshift range $0.8<z<2.2$ where the ISW signal is expected to fade in $\Lambda$CDM, due to a weakening dark energy component, and eventually become consistent with zero in the matter dominated epoch. On the contrary, alternative cosmological models, proposed to explain the excess low-$z$ ISW signals, predicted a sign-change in the ISW effect at $z\approx1.5$ due to the possible growth of large-scale potentials that is absent in the standard model. To discriminate, we estimated the high-$z$ $\Lambda$CDM ISW signal using the Millennium XXL mock catalogue, and compared it to our measurements from about 800 supervoids identified in the eBOSS DR16 quasar catalogue. At $0.8<z<1.2$, we found an excess ISW signal with $A_\mathrm{ISW}\approx3.6\pm2.1$ amplitude. The signal is then consistent with the $\Lambda$CDM expectation ($A_\mathrm{ISW}=1$) at $1.2<z<1.5$ where the standard and alternative models predict similar amplitudes. Most interestingly, we also observed an \emph{opposite-sign} ISW signal at $1.5<z<2.2$ that is in $2.7\sigma$ tension with the $\Lambda$CDM prediction. Taken at face value, these recurring hints for ISW anomalies suggest an alternative growth rate of structure in low-density environments at $\sim100~\mpc$ scales.
\end{abstract}

\section{Introduction}

An over-arching goal of cosmology is the reconstruction of how the minuscule cosmic microwave background (CMB) anisotropies grow to the intricate cosmic web that we observe from galaxy surveys. The key challenge in this process is the deeper understanding of the elusive dark energy component in the consensus $\Lambda$-Cold Dark Matter ($\Lambda$CDM) model, which suppresses the growth rate of structure and speeds up the cosmic expansion rate at low redshifts.

Among the most direct observational tests of dark energy are the weak secondary CMB anisotropies generated by the evolving low-$z$ cosmic web. The dominant signal comes from the late-time integrated Sachs-Wolfe effect \citep[ISW,][]{Sachs1967} at linear scales, while subdominant contributions from the non-linear Rees-Sciama effect \citep[RS,][]{Rees1968} remain at the $\sim10\%$ level compared to the ISW term \citep[see e.g.][]{Cai2010}. 

The linear ISW temperature shift along direction $\hat{\mathbf{n}}$ can be calculated from the time-dependent gravitational potential $\dot{\Phi}\neq0$ based on the line-of-sight integral

\begin{equation}
\label{eq:ISW_definition}
\frac{\Delta T_\rmn{ISW}}{\overline{T}}(\boldsymbol{\hat n}) = 2\int_0^{z_\rmn{LS}} \frac{a}{H(z)}\dot\Phi\left(\boldsymbol{\hat n},\chi(z)\right)\,\rmn{d}z\;,
\end{equation}
with scale factor $a=1/(1+z)$, Hubble parameter $H(z)$, and co-moving distance $\chi(z)$, extending to the redshift of last scattering, $z_\mathrm{LS}$. In the linear growth approximation, density perturbations ($\delta$) grow as $\dot\delta = \dot D\delta$, where $D(z)$ is the linear growth factor. Combined with the Poisson equation for the $\Phi$ potential one can obtain the following ISW formula:
\begin{equation}
\label{eq:ISW_definition2}
\frac{\Delta T_\rmn{ISW}}{\overline{T}}(\boldsymbol{\hat n}) = -2\int_0^{z_\rmn{LS}} a\left(1-f(z)\right)\Phi\left(\boldsymbol{\hat n},z\right)\,\rmn{d}z\;,
\end{equation}
where $f= \mathrm{d}\ln D /\mathrm{d}\ln a$ is the linear growth rate of structure.

Throughout the matter-dominated epoch in $\Lambda$CDM at $z\gtrsim1.5$, the delicate balance of cosmological expansion rate and the growth of structure virtually guarantees constant gravitational potentials ($f\approx1$, $\dot{\Phi}\approx0$) generally in the linear regime of shallow ($|\bar{\delta}| \lesssim 0.3$) density fluctuations averaged over $R\gtrsim 100~\mpc$ scales. CMB photons may traverse hills and wells in the gravitational potential, but their temperatures are not altered as long as the underlying potentials themselves do not change ($\Delta T_\mathrm{ISW}\approx0$). 

However, at low redshifts ($z \lesssim 1.5$) the balance is broken due to the emerging dominance of the dark energy component and its extra space-stretching effects (i.e. sub-critical matter density, $\Omega_{m}<1$). Large-scale potentials decay ($\dot{\Phi}<0$) which slightly changes the energies of photons traversing extended matter density perturbations in the low-$z$ Universe at $\mathrm{\sim 100~\mpc}$ scales \citep[see e.g.][]{Cai2010}. 

The late-time $\Delta T_\mathrm{ISW}$ signal is however calculated to be at the $\mu K$-level, which presents observational challenges. The large uncertainty is due to the primary CMB temperature signal that represents an uncorrelated noise term for such secondary anisotropies. 

A common method to detect the ISW signal has been the measurement of the projected 2-point cross-correlation function of matter density fluctuations and CMB temperature anisotropies \citep[see e.g.][and references therein]{Fosalba2003,ho,gian,Stolzner2018}. However, focusing on the most extreme environments, where most of the signal is generated, we also expect that the ISW effect is accessible by cross-correlating CMB temperature maps with individually identified $\mathrm{\sim 100~\mpc}$ scale galaxy \emph{super-structures}. 

\subsection{ISW anomalies}

In the $\Lambda$CDM model, CMB photons gain net energy traversing superclusters because the potential well is shallower on exit than on entry ($\Delta T_\mathrm{ISW} > 0$). In contrast, CMB photons lose energy in large negative density fluctuations, or supervoids ($\Delta T_\mathrm{ISW} < 0$).

By stacking the CMB map on the positions of hundreds of super-structures, the ISW signal emerges as noise fluctuations cancel. In turn, the fine details of the measured ISW imprint can constrain the physical properties of dark energy in an alternative way \citep[see e.g.][]{Nadathur2012,CaiEtAl2014,Kovacs2018,Adamek2020}.  

An important aspect is that the measured amplitude of the ISW signal ($A_\mathrm{ISW}\equiv \Delta T_\mathrm{obs}/\Delta T_\mathrm{\Lambda CDM}$) is often significantly higher than expected in the concordance model ($A_\mathrm{ISW}=1$). 
Such excess ISW signals were first found by \cite{Granett2008} using luminous red galaxies (LRG) from the Sloan Digital Sky Survey (SDSS) data set. It was then confirmed by several follow-up measurements and simulation analyses that the observed signal from the super-structures is about $A_{\rm ISW}\approx5$ times higher than expected from the $\Lambda$CDM model \citep[see e.g.][]{Nadathur2012,Flender2013,Hernandez2013,Aiola}. Overall, these ISW results are considered anomalous because projected 2-point correlation analyses do not find significant excess ISW signals \citep[see e.g.][]{PlanckISW2015,Hang20212pt}. 

An influential development was the ``re-mapping'' of the SDSS super-structures using more accurate spectroscopic redshifts from the Baryon Acoustic Oscillations Survey (BOSS). Yet, the results were mixed. The anomalous ISW signals were re-detected if the merging of smaller voids into larger encompassing under-densities was allowed in the void finding process \citep{Cai2017,Kovacs2018}. In contrast, no significant excess signals have been reported from the same data set when using matched filters and definitions without void merging \citep[][]{NadathurCrittenden2016}. 

Then, \cite{Kovacs2016} used photo-$z$ catalogues of LRGs from the Dark Energy Survey Year-1 (DES Y1) data set and reported an excess signal; similar to the original observation from SDSS data by \cite{Granett2008}. This analysis was extended to the DES Year-3 data set and the excess ISW signals were confirmed \citep{Kovacs2019}. 

These findings were crucial, because they independently detected ISW anomalies using a \emph{different} part of the sky. In combination with the BOSS results using similarly defined supervoids \citep{Kovacs2018}, the ISW amplitude from BOSS and DES Y3 data is $A_\mathrm{ISW}\approx5.2\pm1.6$ in the $0.2<z<0.9$ redshift range and its origin remains unexplained.

\subsection{Alternative hypotheses: modified gravity?}

The prevailing view has been that the excess ISW signals from supervoids cannot be explained neither by changing the parameters of the $\Lambda$CDM model (given strongly bounding \emph{other} constraints), nor by considering its typical alternatives such as $w$CDM models \citep[see e.g.][]{Nadathur2012,Adamek2020}. Therefore, the ISW tensions might be rare ($\sim3-4\sigma$) statistical fluctuations, or some outside-the-box solution is required to accommodate them, possibly considering growth and expansion histories that are in violation of general relativity; there are no truly viable candidates.

In the domain of \emph{toy-models} for modified gravity scenarios, a proposed explanation for the stronger-than-expected ISW imprints is the AvERA (Average Expansion Rate Approximation) approach \citep{Racz2017}. It is a minimally modified N-body simulation algorithm that uses the separate universe hypothesis to construct an approximation of the emerging curvature models \citep[see e.g.][]{Rasanen2011}. The local expansion rate is calculated on a grid from the Friedmann equations using the \emph{local} matter density, and the expansion rate is averaged over the volume to estimate the zeroth order expansion rate of the simulation box. This treatment of inhomogeneities results in slightly different $H(z)$ expansion and $D(z)$ growth histories compared to a $\Lambda$CDM evolution, including faster low-$z$ expansion rates and a higher $H_0$ value. 

In particular, the enhanced ISW signals from super-structures may be accommodated as a consequence of an even more suppressed growth rate $f(z)$ than in $\Lambda$CDM at $z\lesssim 1.5$, at least at the largest scales \citep[see][for further details]{Beck2018, Kovacs2020}. We note that such qualitative changes in the cosmological model are required by the ISW tensions which, if true, would require radical solutions. The AvERA concept is an alternative to the $\Lambda$CDM evolution. While it is not widely accepted as the most likely alternative, it has a significantly more complex expansion history than most extensions of the dark energy paradigm.

On the observational ground, a counter-argument was already presented by \cite{Hang20212pt} who measured the CMB-galaxy 2-point cross-correlation signal ($C_{\rm gT}$) using the Dark Energy Spectroscopic Instrument (DESI) Legacy Survey photo-$z$ data set \citep{Dey2019}. They found that the AvERA model over-predicts the overall $C_{\rm gT}$ signal at the expense of raising the ISW amplitude in super-structures. Thus they claim that the AvERA model cannot be the final solution for ISW anomalies. 

In a subsequent observational analysis of the super-structures detected in the same DESI Legacy Survey photo-$z$ catalogue, \cite{Hang2021} also questioned the validity of the anomalous ISW signals themselves. They reported no clear detection of ISW signals from their supervoid sample, and a marginal signal with no tension from superclusters. 

These new ISW results further complicate the picture, and they certainly warrant further studies and a better understanding of this important problem in cosmology. It will be especially interesting to explore if new models, similar to AvERA in their spirit but more detailed, could account for both the ISW, $H_0$, and ''lensing-is-low`` tensions in a joint framework \citep[see e.g.][]{Riess2019,Heymans2021}.

\subsection{Testing the evolution of the ISW signal}

In this study, we put the $\Lambda$CDM and AvERA models to test in a novel way to potentially exacerbate, or resolve, the existing ISW anomalies. Extending the redshift range of the observations, we identified supervoids in the un-probed $0.8<z<2.2$ range using the eBOSS DR16 QSO catalogue \citep{Ross2020} and then measured their ISW imprint. 

\begin{figure}
\begin{center}
\includegraphics[width=88mm]{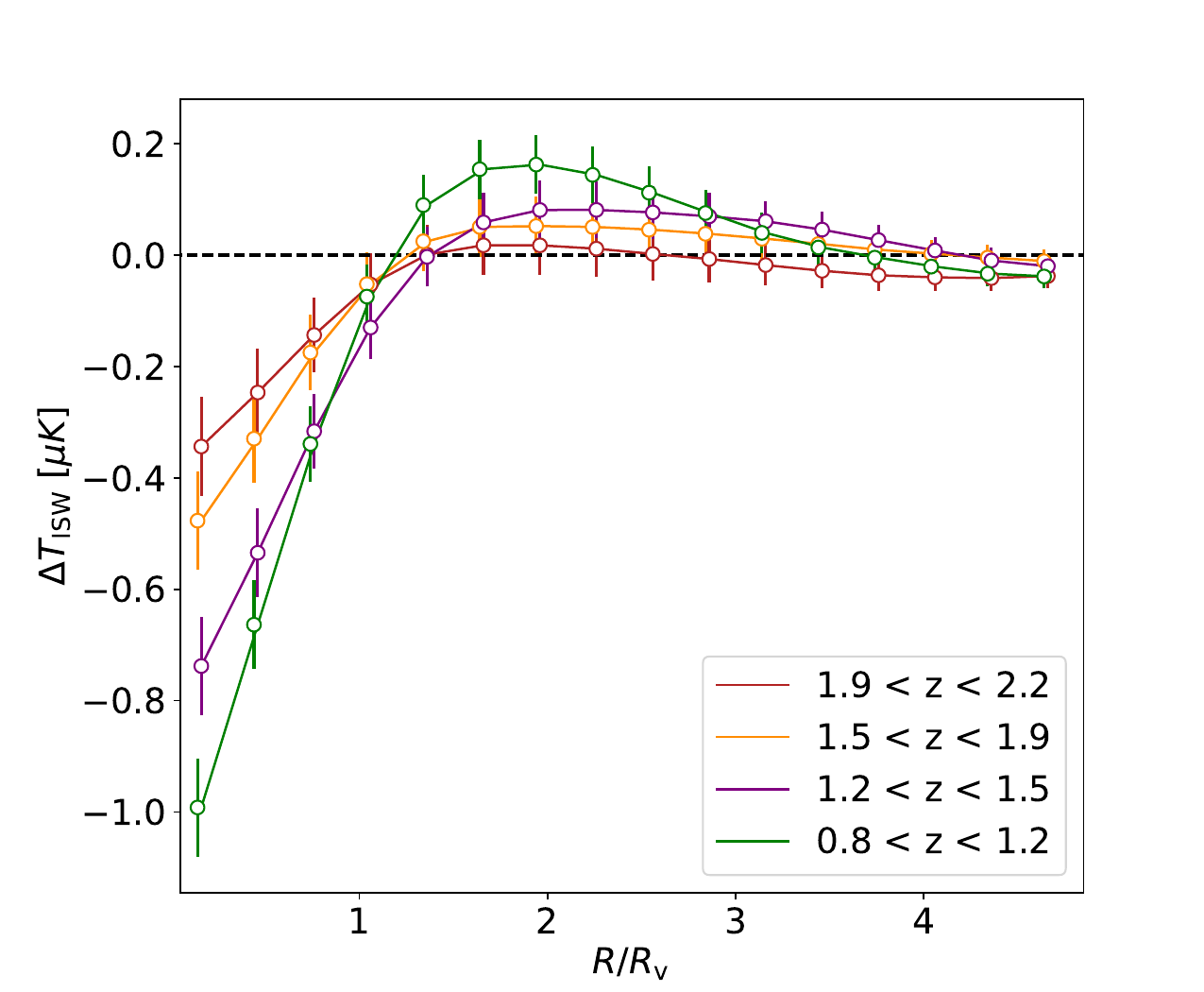}
\caption{\label{fig:figure_1} In the MXXL $\Lambda$CDM simulation, the magnitude of the ISW signal from supervoids (measured by re-scaling to the $R_v$ void radius) decreases towards higher redshift bins. While the shape of the signals can be reliably estimated from the simulated ISW\emph{-only} temperature maps, observational analyses are limited by stronger noise fluctuations from primary CMB anisotropies.}
\end{center}
\end{figure}

This is a key redshift range in the sense that, while the $\Lambda$CDM model predicts a gradually fading signal towards $z\approx2$ (see Figure \ref{fig:figure_1} for a preview of our simulated results), the AvERA model has a characteristically different high-$z$ evolution. In comparison to $\Lambda$CDM, the more suppressed growth rate and faster expansion rate at $z\lesssim1.5$ are compensated by a stronger gravitational growth ($f\gtrsim1$) and a slower expansion rate at $1.5 \lesssim z\lesssim 5$ in the AvERA model \citep[see][for further details]{Beck2018}. 

Therefore, the subtle balance of growth and expansion expected in $\Lambda$CDM at $z\gtrsim1.5$ does not occur in AvERA. We thus formulated a new hypothesis and tested a bold prediction: \emph{if} cosmic expansion and growth are affected by inhomogeneities and this is the true source of the ISW excess signals at low-$z$, \emph{then} there must also be an additional ISW signal observable at $1.5 \lesssim z\lesssim 5$, which is absent in $\Lambda$CDM. In particular, this high-$z$ ISW signal in AvERA is sourced by the \emph{growth} of the potentials ($\dot{\Phi} > 0$), not by their decay like at low-$z$ ($\dot{\Phi} < 0$), and thus it is of opposite sign. This is in stark contrast with $\Lambda$CDM’s fading ISW signal. 

We thus argue that the reconstruction of the ISW signal's evolution in the $0.8<z<2.2$ range can be used to discriminate between these two hypotheses. The first observational study of these high-$z$ ISW signals is the main goal of this paper that is organized as follows. In Section \ref{sec:section_2}, we describe the data sets that are used to measure and model the high-$z$ ISW signals. In Section \ref{sec:section_3}, we provide a summary of our methodology, followed by the presentation of our results in \ref{sec:section_4}. Finally, Section \ref{sec:section_5} contains a discussion and interpretation of our main findings.

\section{Data sets}
\label{sec:section_2}

\subsection{eBOSS quasars}

Our observational analysis of supervoids is based on the Data Release 16 (DR16) QSO sample from the eBOSS survey. The CORE QSO target selection is described by \cite{Myers2015}, using both optical imaging data from SDSS and mid-infrared data from the Wide-field Infrared Survey Explorer survey \citep[WISE,][]{wise}. The complete DR16 QSO catalogue is presented by \cite{Lyke2020} while the QSO clustering catalogue, that we use in this analysis, is described by \cite{Ross2020}. 

The eBOSS DR16 sample contains 343,708 quasars covering a sky area of 4,808 $deg^2$, and spanning the redshift range $0.8<z<2.2$. This sample bridges the gap between BOSS CMASS (contant mass) galaxies at $z<0.7$ and the high redshift quasars at $z>2.2$ that probe the Lyman-$\alpha$ forest fluctuations. This state-of-the-art quasar catalogue has been used for various cosmological analyses \citep[see e.g.][]{Neveux2020,Hou2021,Zhang2021}, including measurements of the growth rate of structure from cosmic voids \citep{Aubert2020}.

\subsection{CMB data}

We estimated the CMB imprint of large-scale structures by using a foreground-cleaned CMB temperature map based on the local-generalized morphological component (LGMCA) method \citep{Bobin2014}. It combines the Wilkinson Microwave Anisotropy Probe 9-year data set \citep[WMAP9,][]{bennett2012} and the \emph{Planck} data products \citep{Planck2020_1}. 

At the scales of our interest, this temperature map provides sufficiently accurate results and it guarantees very low foreground contamination. We also performed tests using the \emph{Planck} 2018 temperature map and our results were fully consistent with this fiducial choice.

\subsection{QSO mock catalogue}

The planned cross-correlations require a catalogue of supervoids and a reconstructed ISW map from the same simulation. Our analysis was based on the Millennium-XXL (MXXL) dark matter only $\Lambda$CDM N-body simulation by \cite{Angulo2012}. The MXXL is an upgraded version of the earlier Millennium run \citep{Springel2005}, covering a co-moving volume of ($3h^{-1}$ Gpc)$^{3}$ with $6720^3$ particles of mass $8.456 \times 10^9 \, M_\odot$. It adopted a cosmology consistent with the WMAP-1 mission results \citep{Spergel2003}.

In particular, we utilized a mock QSO catalogue, based on the halo occupation distribution framework (HOD), selected from the the publicly available full sky MXXL halo light-cone catalogue by \cite{Smith2017}. This mock catalogue covers the $0.8<z<2.2$ redshift range of the eBOSS DR16 QSO data set used in our real-world measurements. 

In the construction of the QSO mock catalogue from the MXXL halos, we followed the HOD prescription presented by \cite{Smith2020}. We note that their estimation of the most realistic HOD parameters to match the eBOSS quasar sample was based primarily on the OuterRim N-body simulation \citep{Heitmann2019}. That mock catalogue has different cosmological parameters compared to MXXL, but the expected differences are not significant for our ISW analyses. 
As discussed by \cite{Smith2020}, the HOD approach describes the average number of central and satellite QSOs residing in halos as a function of halo mass, $M$. The total number of QSOs in a dark matter halo is the sum of the central and satellite quasars, expressed as
\begin{equation}
\langle N_\mathrm{tot}(M) \rangle = \langle N_\mathrm{cen}(M) \rangle + \langle N_\mathrm{sat}(M) \rangle.
\end{equation}

We note that the probability of finding more than one QSO within the same dark matter halo is low because quasars are rare tracers of the underlying density field. Formally, the probability that a halo contains a central QSO is given by the smooth step function
\begin{equation}
\langle N_\mathrm{cen}(M) \rangle = \tau \frac{1}{2} \left[1 + \mathrm{erf} \left( \frac{\log M - \log M_\mathrm{cen}}{\log \sigma_\mathrm{m}} \right) \right],
\end{equation}
where the position of this step is set by $M_\mathrm{cen}$. A halo with mass $M \ll M_\mathrm{cen}$ hosts no central quasar,
which then transitions to a $\tau$ probability for $M \gg M_\mathrm{cen}$ halos. The quasar duty cycle $\tau$ takes into account that not all central black holes are active. It is defined as the fraction of halos which host an active central galaxy, setting the height of the step function, and the width of the transition is set by the parameter $\log \sigma_\mathrm{m}$ \cite[see][for details]{Smith2020}. 

This form of HOD is similar to what is used for galaxies \citep[][]{Tinker2012}, with the addition of the duty cycle parameter, and it is motivated by the expectation that the brightest QSOs occupy the most massive halos.

The number of satellite quasars in each halo is Poisson distributed, with a mean value given by a power law,
\begin{equation}
\langle N_\mathrm{sat}(M) \rangle = \left( \frac{M}{M_\mathrm{sat}} \right)^{\alpha_\mathrm{sat}} \exp \left( - \frac{M_\mathrm{cut}}{M} \right),
\label{eq:hod_satellite_power_law}
\end{equation}
with $\alpha_\mathrm{sat}$ as the slope, $M_\mathrm{sat}$ 
as a normalisation, and $M_\mathrm{cut}$ to apply a cutoff at low masses. These satellite QSOs are randomly positioned in the halo following a Navarro-Frenk-White \citep[NFW,][]{NFW1996} profile.

We followed the ``HOD0'' prescription presented by \cite{Smith2020} which showed the most realistic description of the eBOSS DR16 QSOs. The MXXL halo light-cone catalogue was populated with QSOs using the following parameters: $f_\mathrm{sat}$=0.19, $\tau$=0.012, $\log M_\mathrm{cen}$=12.13, $\log \sigma_M$=0.2, $\log M_\mathrm{sat}$=15.29, $\log M_\mathrm{cut}$=11.61, and $\alpha_\mathrm{sat}$=1.0. 
The corresponding average tracer density $\bar{n}\approx2\times10^{-5}h^{3} \mathrm{Mpc^{-3}}$ is comparable to that of the eBOSS DR16 quasars. Given the large-scale nature of our supervoid analysis, possible small differences in observed and simulated tracer density are not expected to significantly affect our ISW results.

\subsection{Simulated ISW map}

We used the publicly available ISW map reconstruction code by \cite{Beck2018} to produce an ISW map from the MXXL simulation. The same map was used in a previous analysis of MXXL supervoids by \cite{Kovacs2020}. 

We also followed the simulation analysis presented by \cite{Kovacs2019} who reported that large-scale modes add extra noise to the stacked profile and potentially introduce biases in the measured ISW profiles if measured in smaller patches. We therefore removed the contributions from the largest modes with multipoles $2\leq \ell \leq10$.

\section{Methods}
\label{sec:section_3}

In general, cosmic voids are highly hierarchical objects in the cosmic web with two main classes. Voids-in-clouds tend to be surrounded by an over-dense environment, while voids-in-voids, or supervoids, consist of several sub-voids \citep[see e.g.][]{Sheth2004,Lares2017}. 

Such large-scale supervoid structures are of high interest in ISW measurements, as they are expected to account for most of the observable signal. In our analysis, we focused on the identification of such $R\gtrsim100~\mpc$ supervoids, and statistically measured their ISW imprint in the CMB.

\subsection{Supervoid identification}

While various algorithms exist to define cosmic voids, we used the so-called 2D void finding algorithm \citep{Sanchez2016,Davies2021}. The heart of the method is a restriction to tomographic slices of galaxy data, and analyses of the projected density field around void centre candidates defined by minima in the smoothed density field. 

The algorithm includes measurements of galaxy density in annuli about void centre candidates until the mean density is reached, which in turn defines the void radius. Large samples of 2D voids have been used in previous DES void lensing and ISW measurements, showing robust signals from both observed data and simulations \citep[e.g.][]{Vielzeuf2019,Kovacs2019,Fang2019,Kovacs2022b}.

A free parameter in the 2D void finding process is the thickness of the tomographic slices. It was found that an $s\approx100~\mpc$ line-of-sight slicing effectively leads to the detection of independent, and individually significant under-densities \citep{Sanchez2016,Kovacs2019}. We thus sliced the MXXL mock and the eBOSS quasar data set into 18 shells of $100~\mpc$ thickness at $0.8<z<2.2$. 

Another void finder parameter is the Gaussian smoothing scale applied to the tracer density map to define the density minima, which also controls the merging of smaller voids into larger supervoids. In practical terms, while for example a $\sigma=20~\mpc$ smoothing allows one to detect more voids, with $\sigma=50~\mpc$ smoothing a higher number of extended $R_{\rm v}\gtrsim100$ $\mpc$ supervoids are expected in the catalogues as a result of void merging. 

For ISW measurements, such a sample of large voids is beneficial since they carry most of the signal. They better trace the large-scale fluctuations in the gravitational potential which naturally varies on larger scales than the galaxy density field. Motivated also by the large mean inter-tracer separation of quasars, we therefore followed \cite{Kovacs2019} and used $\sigma=50~\mpc$ as a smoothing parameter in our MXXL and eBOSS analyses to detect supervoids.

A third parameter is the minimum central under-density that is considered as a void centre ($\delta_{c}$, measured in the innermost $25\%$ region of voids). We again followed \cite{Kovacs2019} and selected supervoids with $\delta_{c}<-0.3$. 

As a possible consequence of the low QSO tracer density \citep[see e.g.][]{Hawken2020}, we observed that a small fraction of the voids appear completely empty in their centre ($\delta_{c}\approx-1.0$). While these void candidates are smaller than the typical $R\gtrsim100~\mpc$ supervoids in our eBOSS sample, such strong perturbations are not generally expected at such large scales ($R\sim10~\mpc$) in the early stages of the growth of density perturbations. These properties suggest that these voids might potentially be spurious. Nevertheless, their ISW imprints are expected to be low, and they may also be affected by non-linear effects (even if they are real matter under-densities) which are not captured by our linear ISW maps used for modelling in MXXL. For our fiducial analysis, we thus removed these voids from our eBOSS and MXXL samples, but confirmed that they do not change the main conclusions if included in the analysis.

To create a binary \texttt{HEALPix} \citep{healpix} mask for the void finder, we constructed an eBOSS survey mask from the QSO catalogue following a similar eBOSS void analysis by \cite{Aubert2020}. In our MXXL analysis, a full sky mock catalogue was available but we split that into octants to more faithfully model the eBOSS void identification process that makes use of a 4,808 $deg^2$ sky area.

\subsection{The eBOSS and MXXL supervoid samples}

With the above methodology, we identified 8,609 supervoids of radii $R_\mathrm{v}\gtrsim100$ $\mpc$ at redshifts $0.8<z<2.2$ by combining the results from the 8 octants in the full-sky MXXL mock. In eBOSS, which covers approximately $10\%$ of the sky, we identified 838 supervoids (about $10\%$ of the number of supervoids in the MXXL full-sky map).

We then compared the average properties of supervoids in the eBOSS and MXXL catalogues and found great agreement. The maximum supervoid radius in both samples is about $R_\mathrm{v}^{max}\approx350~\mpc$, while the mean radius is slightly larger for the eBOSS sample ($\bar{R_\mathrm{v}}\approx197~\mpc$) compared to the MXXL mock ($\bar{R_\mathrm{v}}\approx174~\mpc$). 

We also found that the eBOSS supervoids are on average about $10\%$ deeper in their central regions ($\bar{\delta}_{c}\approx-0.64$) than supervoids in MXXL ($\bar{\delta}_{c}\approx-0.53$). Transforming these under-density values from galaxy density to matter density given the previusly estimated linear QSO bias factor $b_{Q}\approx2.45\pm0.05$ \citep[][]{Laurent2017}, we got $\bar{\delta}_{c}^{m}\approx-0.26$ for eBOSS supervoids and $\bar{\delta}_{c}^{m}\approx-0.22$ for MXXL supervoids. These relatively shallow structures certainly are rare fluctuations at such large scales in a $\Lambda$CDM model, but they are consistent with the dimensions of the largest known supervoids in combination of their size and emptiness \citep[see e.g.][]{Jeffrey2021,Shimakawa2021}.

We note that neither the mock catalogues and the HOD algorithms nor the eBOSS QSO data set were optimized for the sort of large-scale ISW measurements that we developed in this paper. The observed trends for slightly deeper and larger supervoids in the eBOSS data are consistent with possible imperfections in our HOD modeling, and do not significantly affect our ISW measurements. 

Alternatively, these differences may correspond to genuine physical differences in the evolution of these large-scale structures compared to the baseline $\Lambda$CDM model. This interesting possibility should be better understood in a more detailed analysis of voids, including other void definitions \citep[see e.g.][]{Aubert2020}, and thus we leave these additional tests for future work.

\begin{figure*}
\begin{center}
\includegraphics[width=158mm]{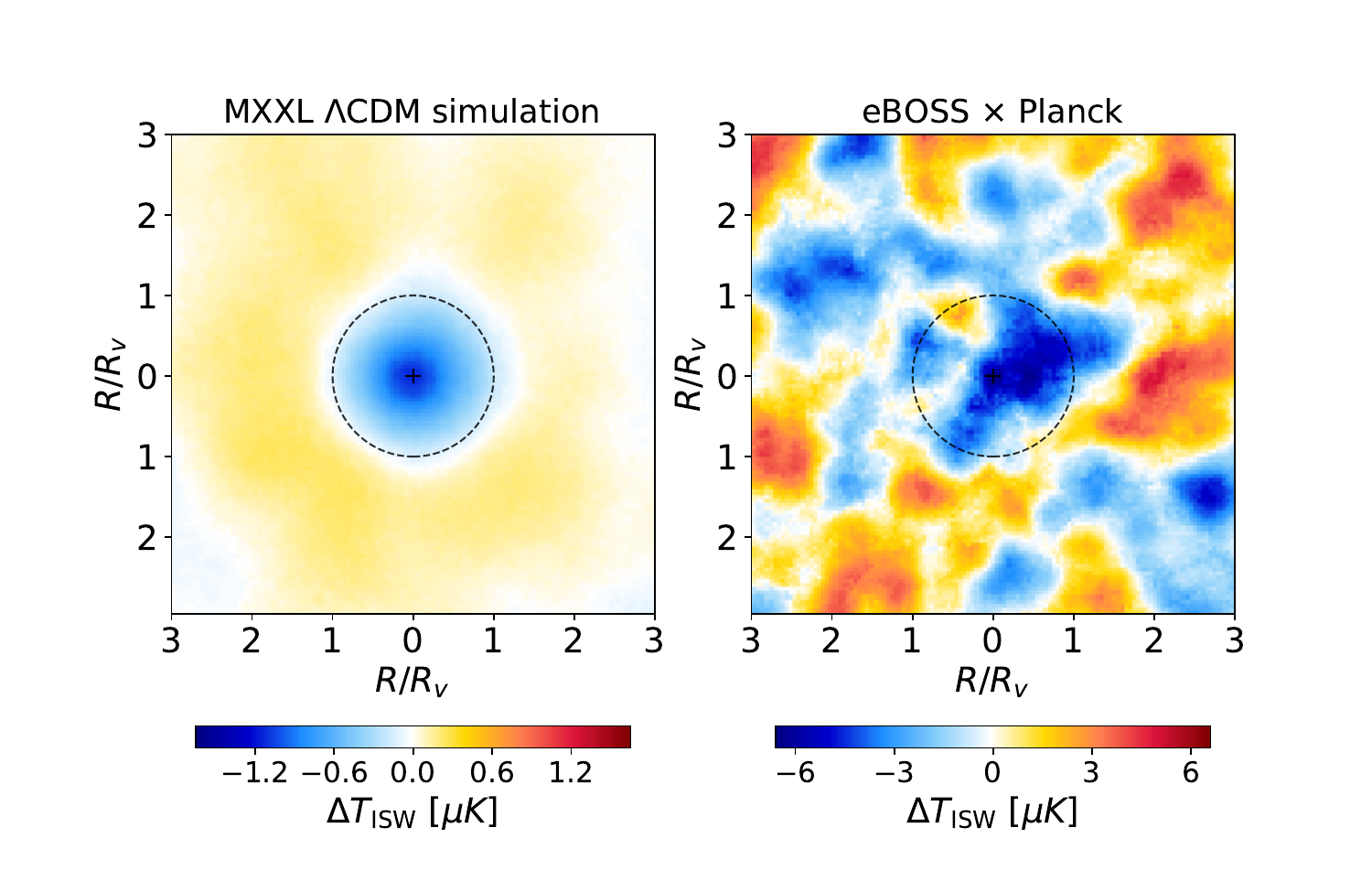}\\
\includegraphics[width=180mm]{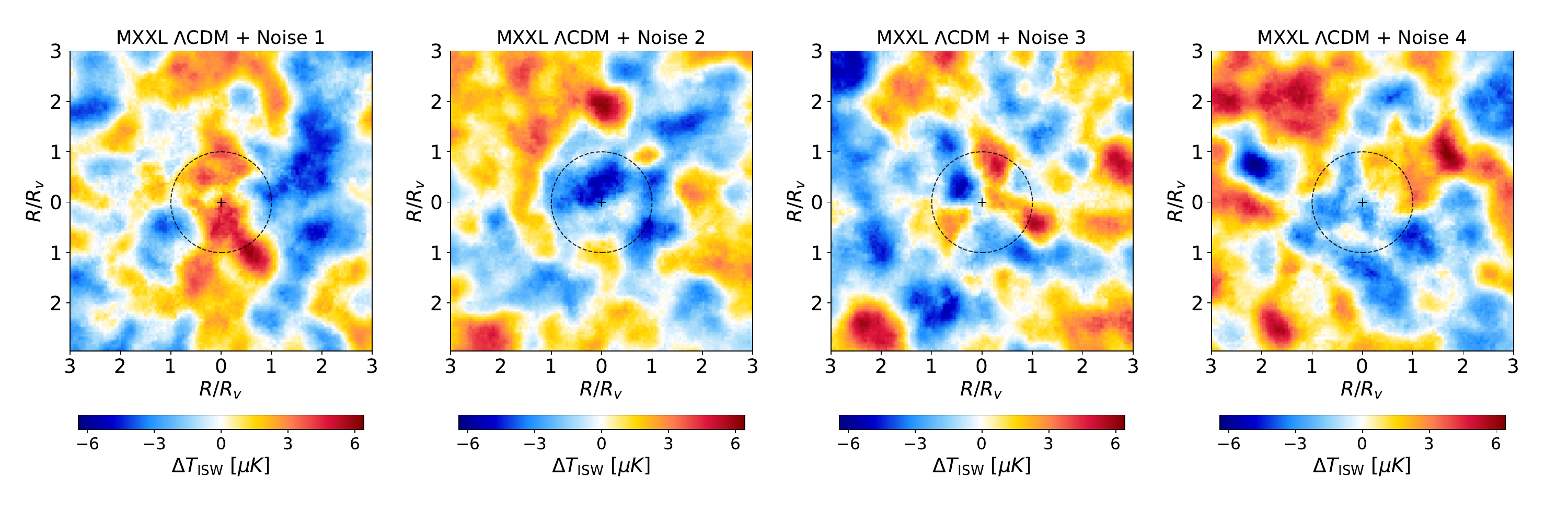}
\caption{\label{fig:figure_2a}\emph{Top:} stacked ISW signals from supervoids at $0.8<z<1.2$ are compared for the ISW-\emph{only} MXXL simulation (left) and the eBOSS QSO data set (right). $R/R_{v}=1$ marks the supervoid radius in re-scaled units, while $R/R_{v}=0$ is the centre where the highest signal is expected. While the eBOSS measurement is affected by significant noise from primary CMB fluctuations, the observational data shows a moderately significant enhanced ISW signal that is similar to previous results from DES and BOSS. \emph{Bottom:} assessing the detectability of the weak $\Lambda$CDM ISW signal at $0.8<z<1.2$ using the eBOSS QSO data set, we added four random realisations of noise from primary CMB temperature fluctuations to the MXXL ISW signal (top-left panel). These realistic examples highlight that the $\Lambda$CDM signal certainly has $S/N<1$ given the important noise from the CMB anisotropies, and patterns like the cold spot seen in the top-right panel can be consistent with chance fluctuations. However, the real-world signal might be stronger and thus more significant. }
\end{center}
\end{figure*}

\begin{figure*}
\begin{center}
\includegraphics[width=155mm]{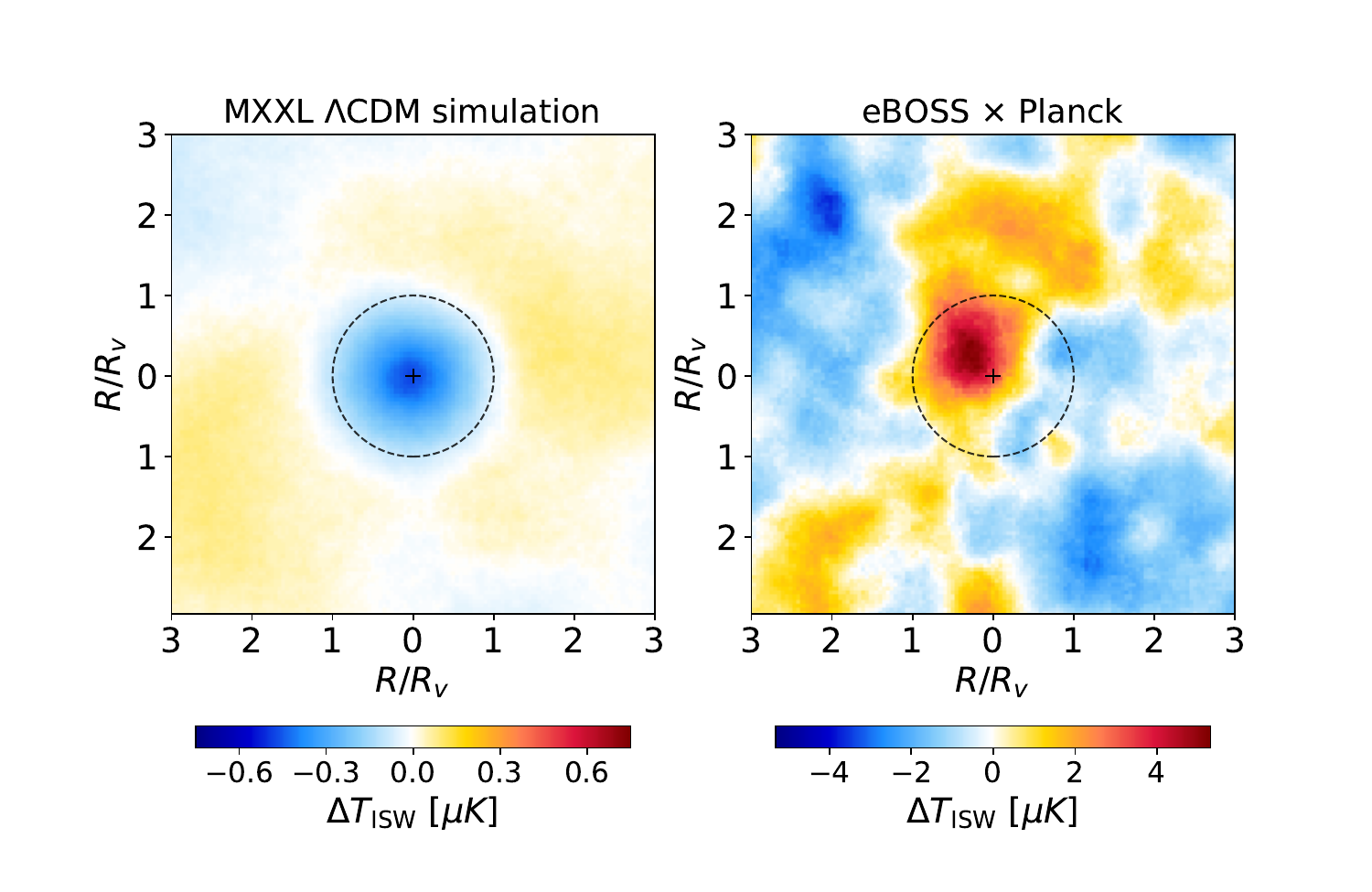}\\
\includegraphics[width=180mm]{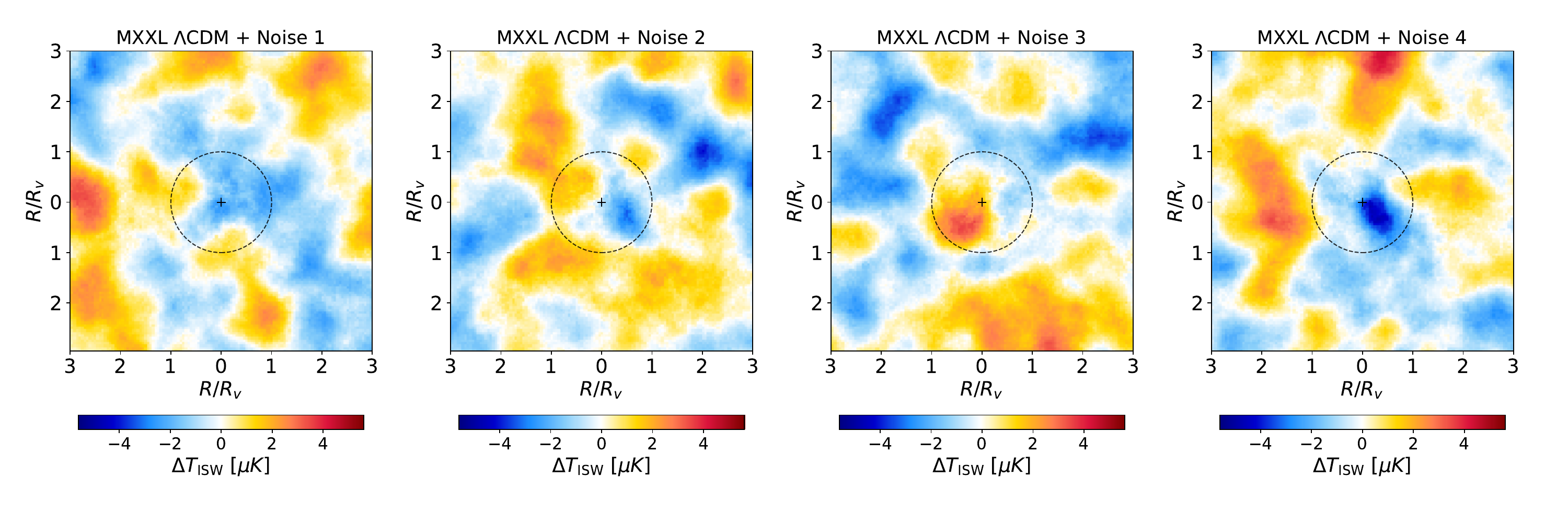}
\caption{\label{fig:figure_2} \emph{Top:} stacked ISW signals from supervoids at $1.5<z<2.2$ are compared for the ISW-\emph{only} MXXL simulation (left) and the eBOSS QSO data set (right). $R/R_{v}=1$ marks the supervoid radius in re-scaled units, while $R/R_{v}=0$ is the centre where the highest signal is expected. We found evidence for a sign-change in the observed ISW imprints at $z\approx1.5$, as predicted by the AvERA model. \emph{Bottom:} assessing the detectability of the $\Lambda$CDM ISW signal at $z>1.5$ using the eBOSS QSO data set, we added four random realisations of noise from primary CMB temperature fluctuations to the MXXL ISW signal (top-left panel). These examples highlight that the $\Lambda$CDM signal certainly has $S/N<1$ given the important noise from the CMB. Consequently, the hot spot pattern seen in the top-right panel can be consistent with chance fluctuations. However, the high-$z$ eBOSS hot spot signal appears to be more significant than the cold spot seen at $0.8<z<1.2$, given the largest number of supervoids in the stacking measurements. We note also that the combination of these two ISW anomalies at different redshifts increases the chance of their validity, either due to physics or systematics.}
\end{center}
\end{figure*}

\begin{figure*}
\begin{center}
\includegraphics[width=86mm]{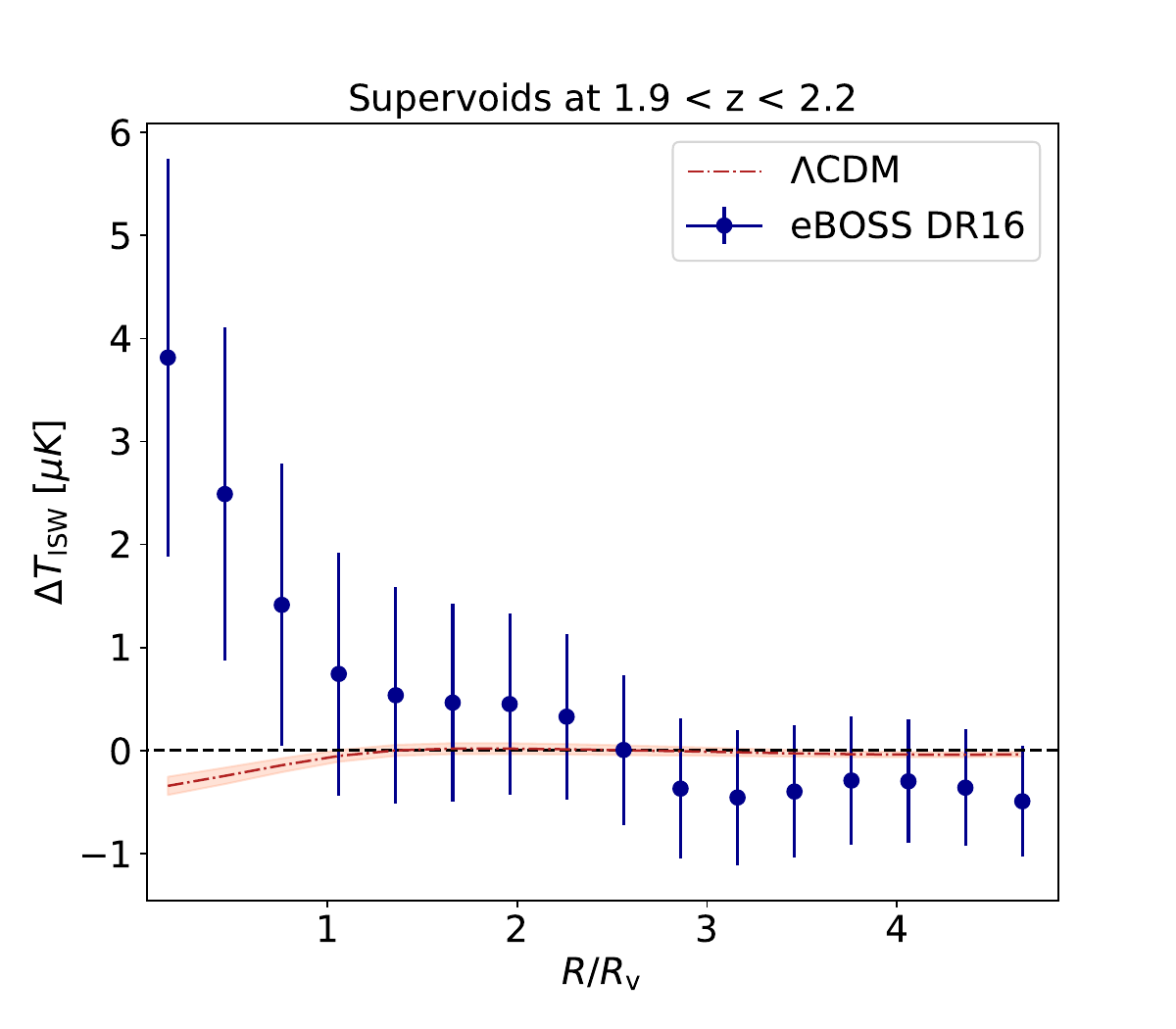}
\includegraphics[width=86mm]{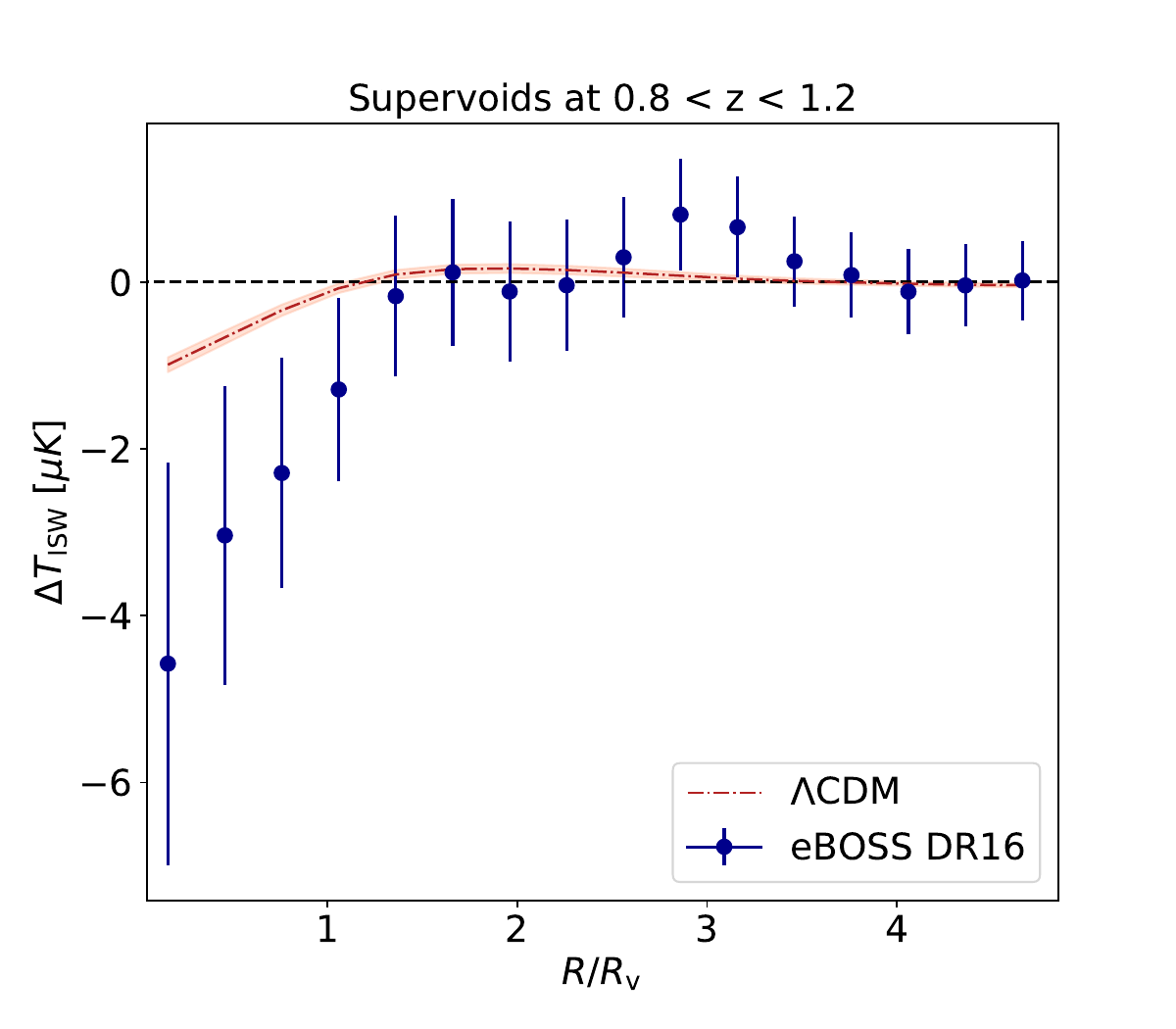}\\
\includegraphics[width=86mm]{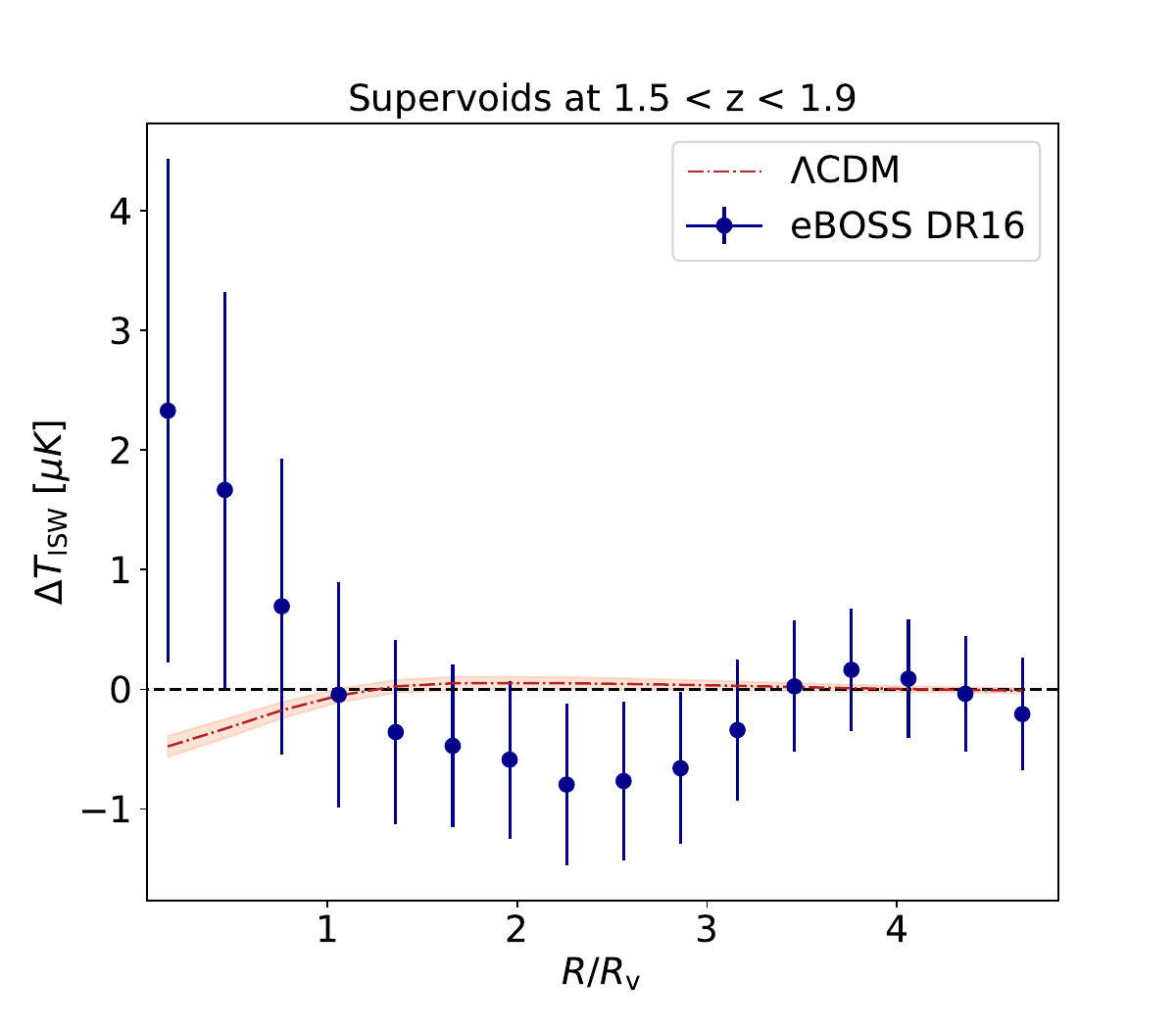}
\includegraphics[width=86mm]{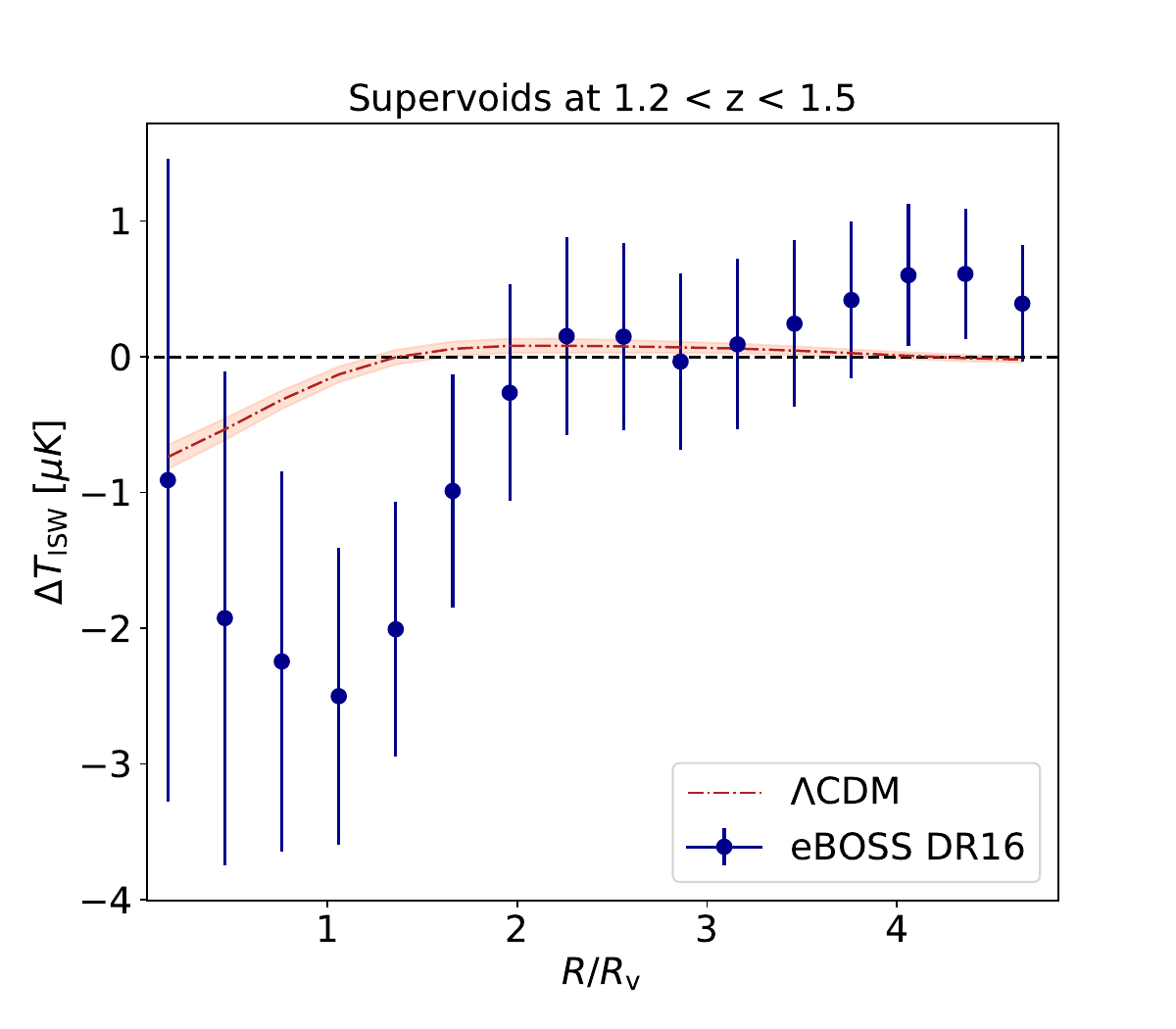}
\caption{\label{fig:figure_3}Measured temperature profiles from stacking analyses in the MXXL simulation and using eBOSS DR16 data. The top-right panel shows our findings of a stronger-than-expected ISW signal at $0.8<z<1.2$ with the sign predicted by the $\Lambda$CDM model. At redshifts $1.2<z<1.5$ (bottom-right), the observed excess ISW signal fades in the centre, but we found traces of an unexpected signal outside the void radius ($R/R_{v}>1$). The two panels on the left show our observation of an \emph{opposite-sign} ISW signal in our two higher-$z$ bins at $1.5<z<1.9$ and at $1.9<z<2.2$. Overall, the observed ISW signals are inconsistent with the $\Lambda$CDM model predictions at all redshifts although the significance of these deviations remains at the moderate $\sim2\sigma$ level.}
\end{center}
\end{figure*}

\subsection{Stacking measurement}

Given the supervoid parameters in the catalogues we constructed, we first cut out square-shaped patches from the CMB temperature maps aligned with supervoid positions using the \texttt{gnomview} projection method of \texttt{HEALPix} \citep{healpix}. In our initial tests, we determined that a $\sigma=1^{\circ}$ Gaussian smoothing applied to the CMB maps is helpful to suppress strong small-scale fluctuations from degree-scale primary CMB anisotropies, and we applied this in our measurements and simulations consistently.

We then stacked the cut-out patches to provide a simple and informative way to statistically study the mean imprints (see Figures \ref{fig:figure_2a} and \ref{fig:figure_2} for examples of stacked ISW images). From the stacked images, we also measured radial ISW profiles in re-scaled radius units using 16 bins of $\Delta (R/R_{v})=0.3$ up to five times the supervoid radius ($R/R_{v}=5$). 

For completeness, we explored the role of the duty cycle of quasars ($\tau$) in our measurements. We analyzed 6 random realisations of our MXXL QSO mock by activating a different subset of halos. At the catalogue level, we detected small changes in the total number of objects with about N$\approx$8,600$\pm$100 supervoids. Concerning the ISW signal amplitude, we found about $10\%$ fluctuations in the stacked imprints. While the overall consistency of these results was good, we decided to take the mean imprint of these 6 realisations as our estimate of the ISW imprints from MXXL supervoids for more accurate results.

We also tested the MXXL ISW signals by applying a simple $\log M$>12.0 halo mass cut which provides a more dense tracer catalogue. Compared to a sub-sampled QSO catalogue, we found about $\sim20\%$ stronger central ISW imprints due to a presumably higher precision to identify the centres of the supervoids where the signal is the strongest. These results confirmed the intuition that future QSO catalogues with higher object density will provide a better chance to measure these signals \citep[see e.g.][]{DESI}.

\section{Results}
\label{sec:section_4}

In order to study the expected redshift evolution of the ISW signal, we decided to split our MXXL and eBOSS supervoid catalogues into the following 4 redshift bins: $0.8<z<1.2$, $1.2<z<1.5$, $1.5<z<1.9$, and $1.9<z<2.2$. This choice results in a fairly equal distribution of the 838 eBOSS supervoids with about 200 of them placed in each redshift bin. A similar split was applied to the 8,609 MXXL supervoids with approximately 2,000 objects in each bin. 

This analysis setup provides sufficient statistical power to explore the expected trends in the data. Importantly, it was also expected to provide new insights about the hypothesized \emph{sign-change} in the ISW signal at $z\approx1.5$.

\subsection{ISW in the MXXL mock}

In our simulations, we observed the expected trend in the evolution of the $\Lambda$CDM ISW amplitude. The amplitude of the signal decreases with increasing redshift, as a result of the transition towards the Einstein-de Sitter-like matter dominated universe with $\dot{\Phi}\approx0$ from about $z\approx2$. 

In Figure \ref{fig:figure_1}, we show the estimated ISW\emph{-only} signals in the 4 redshift bins. We calculated the corresponding ``theoretical'' uncertainties of the full-sky ISW\emph{-only} signals from 500 random stacking measurements. Given the ISW auto power spectra \cite{Beck2018} calculated from the MXXL mock, we generated 500 realisations of ISW\emph{-only} maps using the \texttt{synfast} routine of \texttt{HEALPix}. We then used the MXXL supervoids for stacking measurements on these uncorrelated maps to estimate the uncertainties of the ISW profile reconstruction itself due to fluctuations in the ISW signal map.

\subsection{ISW from eBOSS supervoids}

Next, we measured the imprint of the real-world eBOSS supervoids on the observed CMB temperature anisotropy map. In Figures \ref{fig:figure_2a} and \ref{fig:figure_2}, we compare two stacked images that we created using MXXL and eBOSS supervoids located at $0.8<z<1.2$ and at $1.5<z<2.2$, respectively. While the low-$z$ eBOSS data shows an enhanced cold imprint, the high-$z$ part provides a strong visual impression of qualitatively different ISW imprints with a central cold spot in the $\Lambda$CDM model, and a \emph{hot spot} imprint from the eBOSS data. 

Following a standard approach in ISW measurements \citep[see e.g.][]{Kovacs2019}, the errors of these stacking measurements were estimated based on 500 random CMB map realisations using the \texttt{synfast} routine. We used the CMB angular power spectrum estimated from the \emph{Planck} data set \citep{Planck2018_cosmo} to generate the random CMB maps (including the dominant primary anisotropies), and then stacked the eBOSS supervoids on them. We did not change the supervoid positions via randomisation, in order to keep their overlaps and internal correlations. We then calculated the covariance (C) of our real-world measurements from this ensemble of random measurements.

We note the MXXL $\Lambda$CDM ISW signal does \emph{not} include noise fluctuations from the primary CMB anisotropies in the top panels of Figures \ref{fig:figure_2a} and \ref{fig:figure_2}. We thus created more realistic stacked ISW images using random realisations of primary CMB fluctuations added to the MXXL-based signals, given the eBOSS supervoids in the redshift bins $0.8<z<1.2$ and $1.5<z<2.2$ that we analysed. This comparison allows a more realistic determination of the significance of the hot/cold spot patterns seen aligned with eBOSS supervoids, which in fact seem to be consistent with some random fluctuations. On the other hand, our \emph{prior} knowledge about similar excess ISW signals with an $A_\mathrm{ISW}\approx5.2\pm1.6$ amplitude from DES Y3 and BOSS supervoids \citep{Kovacs2019} at $0.2<z<0.9$ suggests that these high-$z$ ISW excess signals might be real anomalies in the eBOSS and \emph{Planck} data sets.

Given these observed ($\Delta T^\mathrm{o}$) and simulated ($\Delta T^\mathrm{s}$) results, we evaluated a chi-square statistic with 
\begin{equation}
{\chi}^2 = \sum_{ij} (\Delta T_{i}^\rmn{o}-A_{\rm ISW}\Delta T_{i}^\rmn{s} )C_{ij}^{-1} (\Delta T_{j}^\rmn{o}-A_{\rm ISW}\Delta T_{j}^\rmn{s})
\end{equation}
where indices $i,j$ correspond to radial bins measured from the stacked image in a given redshift slice. Under the fair assumption of Gaussian likelihoods, we then looked for the maximum of the $\mathcal{L}\sim$ exp$(-\chi^{2}/2)$ function in each redshift bin to determine the best-fit $A_{\rm ISW}$ amplitude and its uncertainties. This is a standard method in the field and it makes comparisons to other ISW results more straightforward.

\subsection{Main findings}

We further examined the ISW signals in our 4 redshift bins, and present our main results in Figure \ref{fig:figure_3}, which allows an assessment of the consistency of the eBOSS and MXXL results given the uncertainties, without any  $A_\mathrm{ISW}$ fitting analysis. We then fit the $\Lambda$CDM template profile to the eBOSS data with a varying $A_\mathrm{ISW}$ amplitude, and kept the shape of the ISW imprint profiles fixed. 

Here we note that, as the expected  ISW signals goes to \emph{zero} towards higher redshifts in the $\Lambda$CDM model, the determination of the $A_\mathrm{ISW}$ observation-to-simulation ratio parameter becomes increasingly noisy. Nonetheless, Figures \ref{fig:figure_2a}, \ref{fig:figure_2}, and  \ref{fig:figure_3} offer a chance to compare the simulated and observed signals without the ISW amplitude-fitting procedure and visually assess the significance of any deviation between data and simulations in the light of the uncertainties. We found that the visually determined significance from stacked ISW images with a $\Lambda$CDM signal plus random noise realisations added to it (Figures \ref{fig:figure_2a}, \ref{fig:figure_2}), and the errors on radial ISW profiles determined from random measurements (Figure \ref{fig:figure_3}) are fully consistent with each other ($S/N\lesssim2$, see below).

Given the above cautionary notes on the interpretation of our ISW amplitude-fitting results, we made the following observations on the $A_\mathrm{ISW}$ amplitudes determined from simulated MXXL and measured eBOSS ISW profiles:
\begin{itemize}
\item at redshifts $0.8<z<1.2$, we found an excess ISW signal with $A_\mathrm{ISW}\approx3.6\pm2.1$ amplitude. This appears to be consistent with the $A_\mathrm{ISW}\approx5.2\pm1.6$ amplitude constrained from the DES Y3 and BOSS data at $0.2<z<0.9$.

\item the observed eBOSS signal is consistent with the $\Lambda$CDM expectation ($A_\mathrm{ISW}\approx-0.9\pm2.9$) at $1.2<z<1.5$, where the $\Lambda$CDM and AvERA models predict similar amplitudes. The potentially spurious signals that are seen outside the supervoids ($R/R_{v}>1$) do not significantly affect the overall best-fit amplitude, that is also consistent with zero signal.

\item we measured $A_\mathrm{ISW}\approx-4.3\pm3.8$ at $1.5<z<1.9$, where the sign of the ISW signal is expected to start changing in the AvERA model. While the significance of the measured signal is low, the data appears to show characteristic features at the centre where the highest signal is expected.

\item finally, we found $A_\mathrm{ISW}\approx-8.49\pm4.4$ in the fourth redshift bin at $1.9<z<2.2$, which provides further evidence for a sign-change in the ISW amplitude.
\end{itemize} 

\begin{figure}
\begin{center}
\includegraphics[width=85mm]{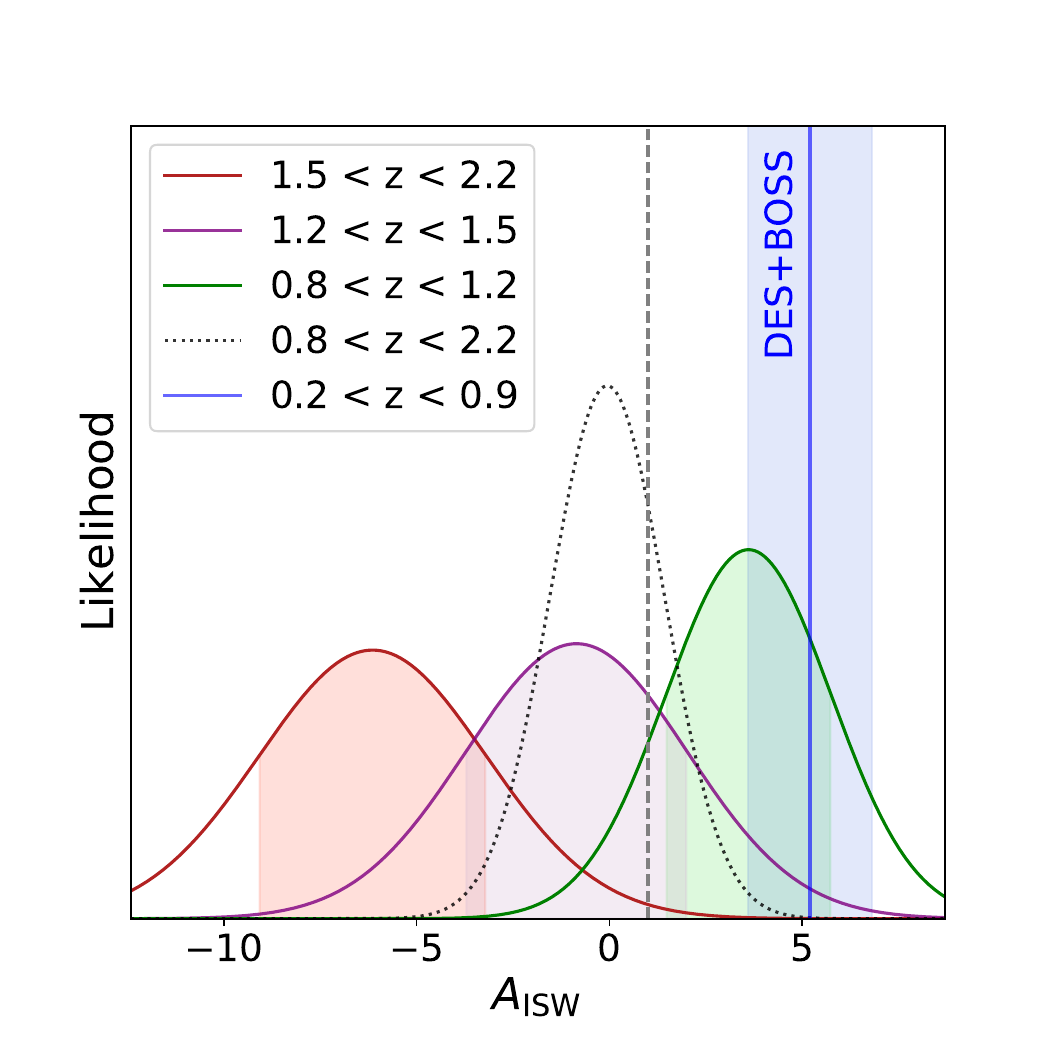}
\caption{\label{fig:figure_4} Likelihood functions of the $\Lambda$CDM model multiplied by an $A_{\rm ISW}$ amplitude in the light of our eBOSS measurements in different redshift bins (shaded regions mark the $A_\mathrm{ISW}^\mathrm{best-fit}\pm1\sigma$ range). The blue band marks the best-fit $A_\mathrm{ISW}$ results from supervoids in previous low-$z$ measurements. The dotted line shows the overall likelihood of the $\Lambda$CDM model \emph{without} redshift binning (consistent with $A_\mathrm{ISW}\approx1$ and also with zero ISW signal).}
\end{center}
\end{figure}

An interesting aspect of our ISW measurement is that the total stacked signal is formally consistent with zero \emph{without} redshift binning, as shown in Figure \ref{fig:figure_4}. The joint  $1.5<z<2.2$ ISW signal from the two high-$z$ bins favours large negative $A_\mathrm{ISW}$ values, the $1.2<z<1.5$ bin is consistent with zero and also the $\Lambda$CDM prediction, while the $0.8<z<1.2$ bin alone constrains a large positive ISW amplitude. In a stacking measurement without binning, the overall imprint is consistent with zero ISW signal due to the cancellation of the high-$z$ and low-$z$ signals which contribute to the mean imprint with different sign. This feature highlights the importance of considering  alternative hypotheses and of executing deeper explorations of the available data.

\begin{figure}
\begin{center}
\includegraphics[width=84mm]{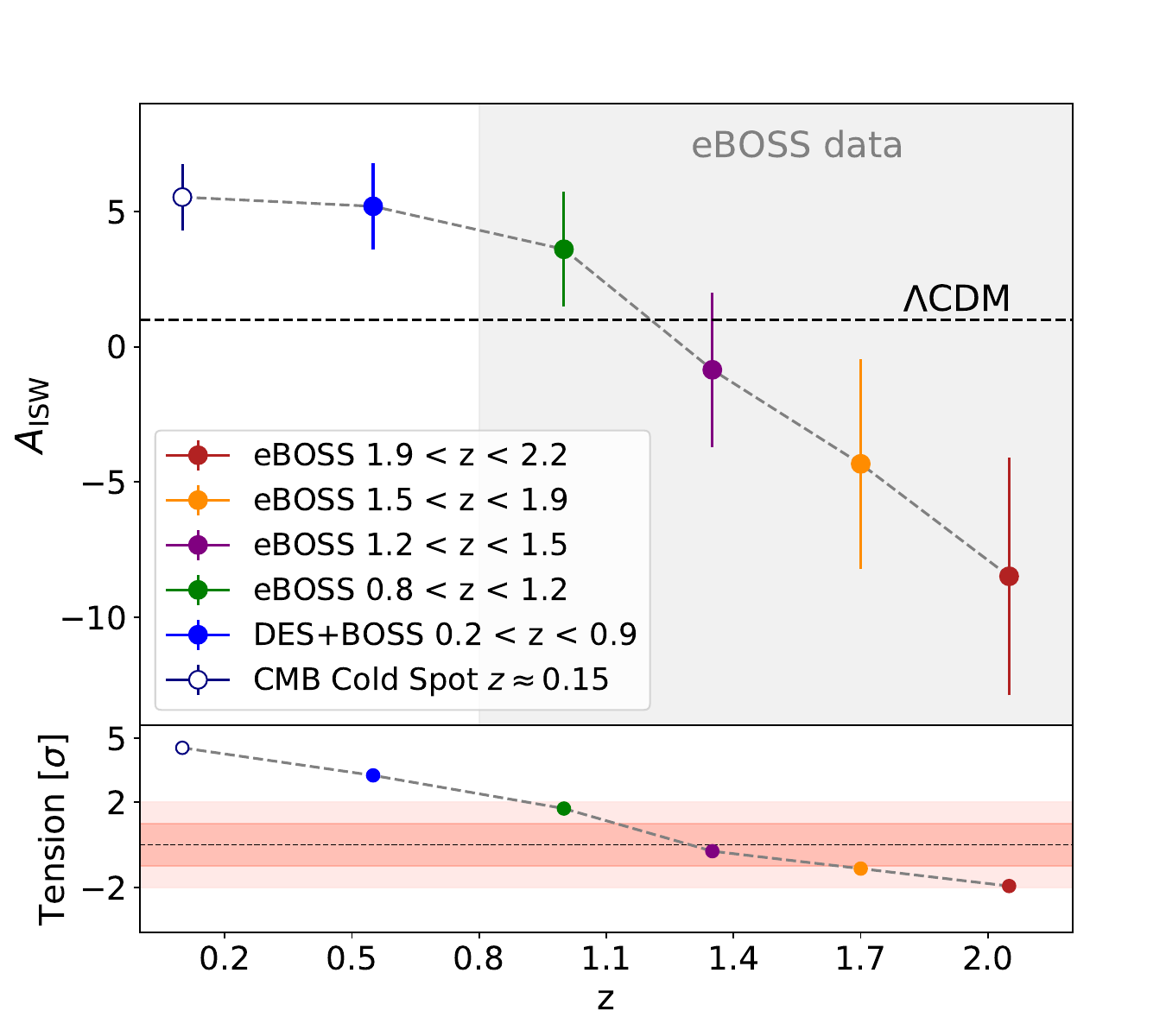}
\caption{\label{fig:figure_6} Measured ISW amplitudes with emerging trends. At $z\lesssim1.5$, multiple observational results point to an enhanced positive ISW amplitude. In contrast, our new eBOSS results showed a large \emph{negative} ISW amplitude at $z\gtrsim1.5$. The bottom panel shows the estimated tension with the baseline $\Lambda$CDM model predictions (shaded bands correspond to $1\sigma$ and $2\sigma$).}
\end{center}
\end{figure}

While the $R\sim100~\mpc$ scales that we probe using the eBOSS supervoid sample suggest that a linear ISW approach is a fair approximation, we also considered the role of possible \emph{second-order} RS effects beyond the linear signal. It is important to note that N-body simulation analyses have shown \citep[see e.g.][]{Cai2010} that about $30\%$ of the central temperature imprint from an extended under-density at $z\approx2$ can be sourced by non-linear RS terms in this limit of fading linear ISW signals. However, the RS effect in the centres of voids acts to \emph{reinforce} the linear ISW signal and, assuming $\Lambda$CDM physics, makes a central cold spot slightly colder. This is contrary to the apparent hot spot ISW signal that we observed in our noisy measurements at $z\gtrsim1.5$.

In Figure \ref{fig:figure_6}, we provide comparisons to existing low-$z$ results from BOSS and DES Y3 data, and visualize the redshift trends in ISW anomalies. For completeness, we also include the formal ISW amplitude enhancement ($A_\mathrm{ISW}\approx5.5$) required to fully explain the CMB \emph{Cold Spot} as an ISW imprint from the Eridanus supervoid it is aligned with \citep[see e.g.][]{SzapudiEtAl2014,KovacsJGB2015,Kovacs2020,Kovacs2022a}. Interestingly, it also follows the same ISW anomaly trend at very low redshifts ($z\approx0.15$).

\subsection{Fluctuations in the expected ISW signal}

While our main analysis is based on a more accurate full-sky estimation of the ISW imprints expected in the $\Lambda$CDM model, we also tested the strength of possible ``cosmic variance'' fluctuations in the expected signal. Given the size of the eBOSS survey footprint and the randomness in QSO detection from the observed volume, we thus measured the stacked cut-sky ISW signals from supervoids in the 8 octants of the MXXL mock using the 6 different realisations of a random QSO activation (48 slightly different patches). 

\begin{figure}
\begin{center}
\includegraphics[width=89mm]{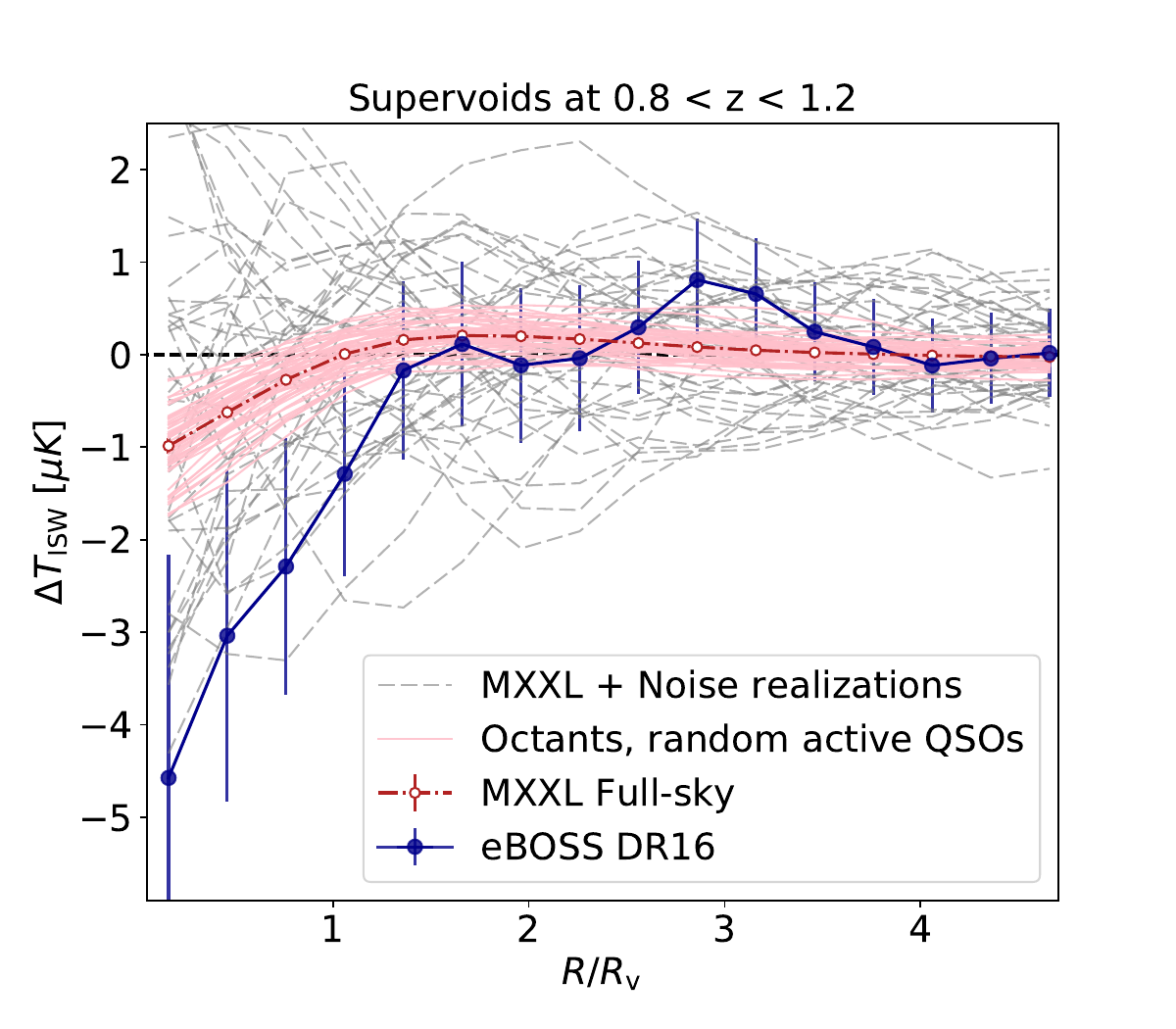}
\caption{\label{fig:figure_5b} A comparison of the anomalous ISW signals from eBOSS data with possible fluctuations (light red) in the expected $\Lambda$CDM imprints from simulations. Measured in different octants and considering 6 different random realisations of the QSO activation in the HOD modeling, the observed excess signal is stronger than the most significant fluctuations in different MXXL mock realisations. We also show another case where the realistic CMB noise from primary anisotropies is also added to the MXXL ISW map (lightgray), showing larger variations and thus better consistency with our eBOSS observations in this redshift range.}
\end{center}
\end{figure}

In Figure \ref{fig:figure_5b}, we show the corresponding results for the $0.8<z<1.2$ redshift bin (other bins show consistent results). In comparison to the full-sky MXXL result, we found that there are considerable variations in the reconstructed locally measured MXXL ISW signals in eBOSS DR16-like observational windows. However, we concluded that even the most extreme fluctuations are insufficient to explain the discrepancy with the observations. 

As an actually dominant noise term, we then also added fluctuations from the primary anisotropies of the CMB to the signal-\emph{only} MXXL ISW map (as in Figures \ref{fig:figure_2a}, \ref{fig:figure_2}). Naturally, the eBOSS results are more consistent with the $\Lambda$CDM model predictions, given such an increased and fully realistic error budget. This visual impression is also consistent with our determination of a mild $1.7\sigma$ hint of an ISW signal with a best-fit $A_\mathrm{ISW}\approx3.6\pm2.1$ amplitude for this redshift range.

We stress, however, that the most interesting aspect of our eBOSS ISW results is not the existence of individual $1-2\sigma$ tensions compared to $\Lambda$CDM expectations in different redshift bins, but an emerging trend for (anti-)correlated deviations from $\Lambda$CDM at low and high redshifts with opposite signs, as shown in Figure \ref{fig:figure_6}.

\subsection{CMB lensing tests}

In the light of the anomalous opposite-sign ISW signals from the $z\gtrsim1.5$ range, we decided to further test the validity of the eBOSS supervoid sample. We stacked the \emph{Planck} CMB \emph{lensing} convergence ($\kappa$) map \citep{Planck2018_cosmo} on the positions of supervoids, by splitting the data into two bins at $z=1.5$. 

\begin{figure}
\begin{center}
\includegraphics[width=86mm]{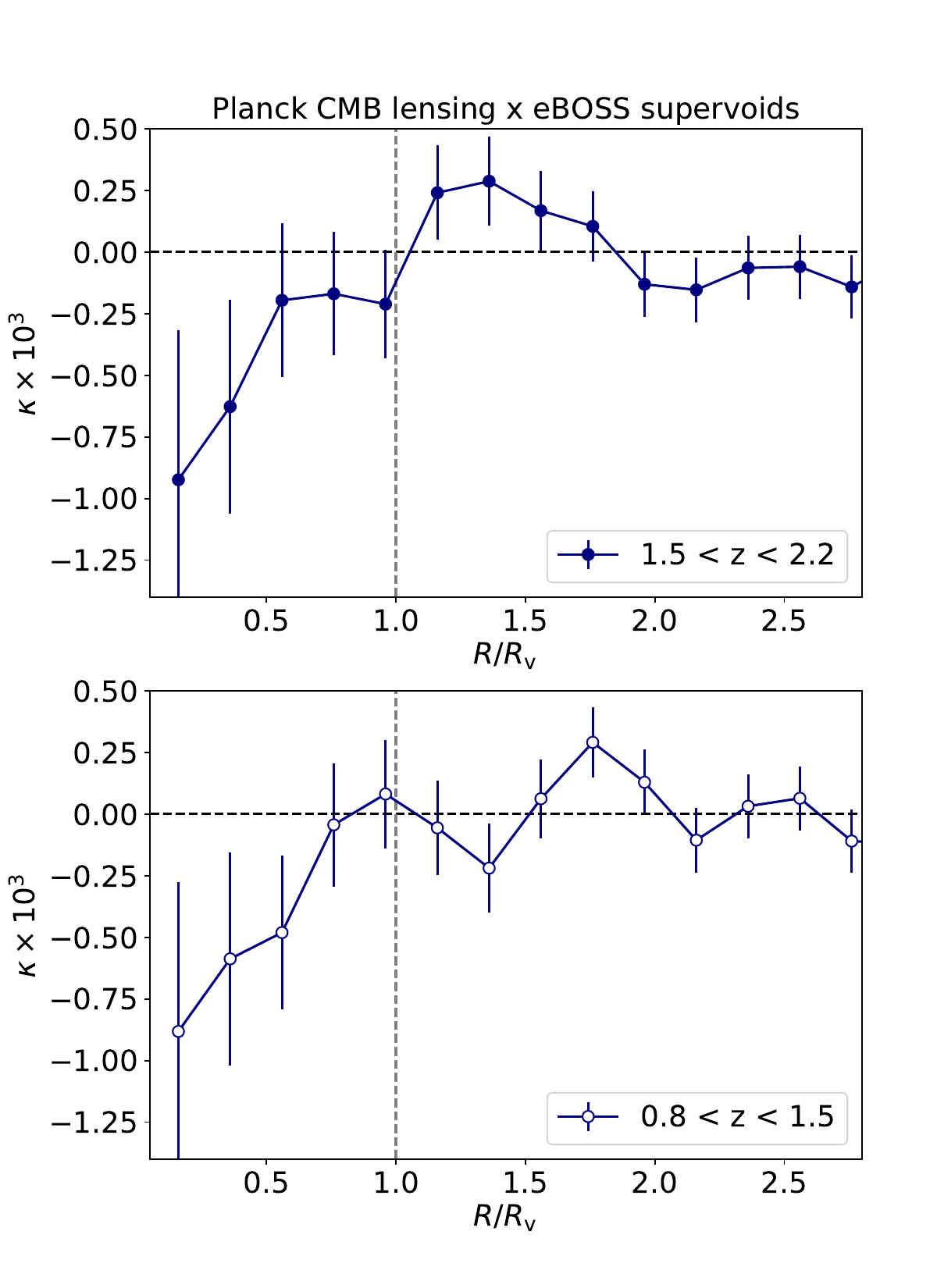}
\caption{\label{fig:figure_5} CMB lensing convergence ($\kappa$) profiles are measured for the high-$z$ (top) and low-$z$ (bottom) parts of the eBOSS supervoid catalogue. Both populations show negative $\kappa$ imprints aligned with the interior of the supervoids, indicating true under-densities.}
\end{center}
\end{figure}

As demonstrated in Figure \ref{fig:figure_5}, we found that both halves of the catalogue show a generally negative $\kappa$ imprint at $R/R_{v}<1$. This finding suggests that despite the evidence for opposite-sign ISW signals from $z>1.5$ eBOSS supervoids, these objects also correspond to genuine under-densities in the cosmic web.

\section{Discussion and Conclusions}
\label{sec:section_5}

Motivated by recently reported anomalies in the ISW signal from the low-$z$ Universe, we extended the redshift range of the relevant observations using the eBOSS DR16 QSO catalogue \citep[][]{Ross2020}. We modeled our measurement with the Millennium XXL simulation \citep[][]{Angulo2012,Smith2017}, and estimated the $\Lambda$CDM ISW signal in the $0.8<z<2.2$ redshift range. We then compared the observed signal from eBOSS supervoids to this $\Lambda$CDM expectation by fitting an $A_{\rm ISW}$ amplitude to the data as a consistency test. These measurements revealed a new ISW anomaly associated with supervoids identified at redshifts higher than before.

Considering possible systematic effects, the cross-correlation nature of our CMB stacking measurements using positions of distant supervoids minimizes the chance of confusing the expected ISW signal with remnant local contamination in the foreground-cleaned CMB data. We also note that the observed redshift dependence of the ISW amplitude, above all the hot spot signals from $z>1.5$ eBOSS supervoids, is inconsistent with a hypothetical contamination from a density-dependent dust emission \citep[see e.g.][for a similar low-$z$ analysis]{Hernandez2013}.

\subsection{An opposite-sign ISW signal}

We also considered an alternative hypothesis. Guided by the nature of the known ISW tensions at $0.2<z<0.9$ and the proposed solution provided by the AvERA model \citep[][]{Racz2017,Beck2018}, we specifically looked for a sign-change in the ISW signal at about $z\approx1.5$. 

Here we note that ISW analyses were not among the key projects in the eBOSS survey, and thus the QSO catalogue was not optimised for such measurements. The expected signal-to-noise is relatively low due to significant noise from the primary CMB fluctuations.

Yet, we did find evidence for such an opposite-sign ISW signal from the eBOSS supervoids which, instead of a cold spot signal, showed a hot spot imprint in the $1.5<z<1.9$ and $1.9<z<2.2$ redshift bins. Combining these two high-$z$ bins, the eBOSS data provided a moderate $2.4\sigma$ hint for an \emph{opposite-sign} ISW imprint at $1.5<z<2.2$. Formally, the actual tension with the $\Lambda$CDM model prediction is $2.7\sigma$ (see Figure \ref{fig:figure_4} for more detailed likelihood analysis results). 

Considering our new eBOSS results and the excess ISW signal from supervoids at $0.2<z<0.9$ detected from BOSS and DES Y3 data \citep[][]{Kovacs2019}, an emerging trend is seen in the data as displayed in Figure \ref{fig:figure_6}. 
At $z\lesssim1.5$, multiple observational results point to an enhanced positive ISW amplitude. Furthermore, our new eBOSS results showed a large \emph{negative} ISW amplitude at $z\gtrsim1.5$.

With additional tests, we confirmed that these $z\gtrsim1.5$ supervoids with the most anomalous hot spot ISW signal are also aligned with negative CMB lensing convergence ($\kappa<0$), indicating genuine under-densities. Moreover, we showed that the discrepancy is not resolved by considering possible ``cosmic variance'' fluctuations in the expected $\Lambda$CDM signal given the eBOSS survey window (see Figure \ref{fig:figure_5b}). 

\subsection{Interpretation $\&$ future prospects}

\begin{figure}
\begin{center}
\includegraphics[width=89mm]{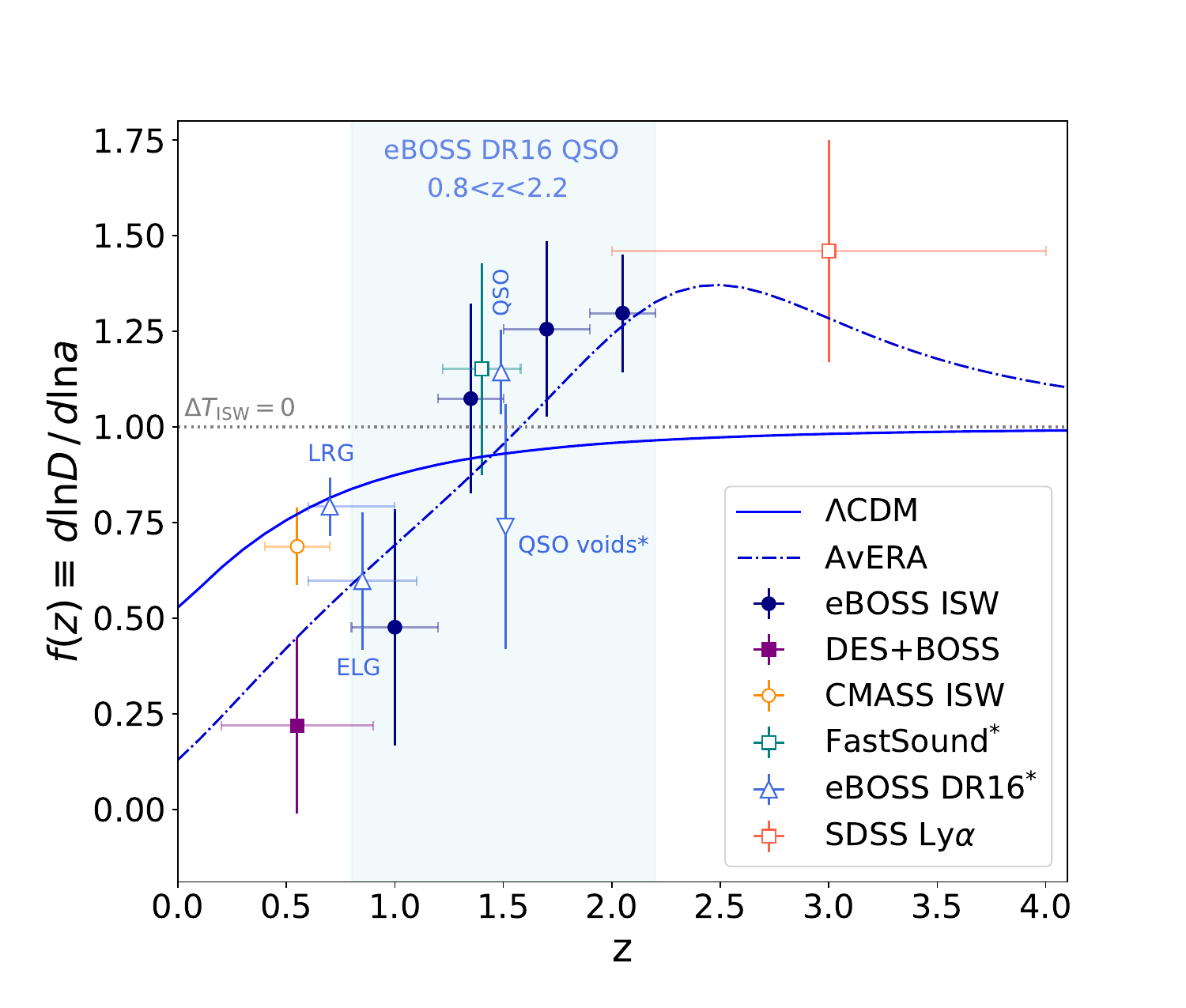}
\caption{\label{fig:figure_7} Growth rate of structure as a function of redshift in the standard $\Lambda$CDM model (assuming a \emph{Planck} 2018 cosmology) and in a variant of the AvERA model \citep[as approximated by][]{Hang20212pt}. Measurements indicated with an asterisk are based on $f\sigma_{8}(z)$ constraints divided by a fiducial \emph{Planck} $\sigma_{8}(z)$ value. If the measured excess amplitudes are formally expressed as re-scaled $\Lambda$CDM growth rate values ($A_{\rm ISW}(z)\times[1-f^{\Lambda CDM}(z)]$), then the ISW anomalies from DES, BOSS, and eBOSS data follow a consistent trend. We note that the ISW amplitude constraint from CMASS data \citep[][]{NadathurCrittenden2016} and eBOSS LRGs do not show significant anomalies. However, eBOSS ELGs, QSOs, and other high-$z$ constraints from the FastSound and SDSS Ly$\alpha$ data are consistent with the results of this paper.}
\end{center}
\end{figure}

Taken at face value, these moderately significant $2-3\sigma$ tensions in the ISW signals in the entire observed redshift range suggest an alternative growth rate of structure; at least in low-density environments at $\sim100~\mpc$ scales. As shown in Figure \ref{fig:figure_7}, the AvERA model appears to provide a framework to interpret these anomalies, if the observed excess ISW amplitudes are interpreted as an enhancement compared to the expected $\Lambda$CDM growth rate of structure with $\Delta T_\mathrm{ISW}(z)\sim [1-f^{obs}(z)]\equiv A_\mathrm{ISW}(z)\times[1-f^{\Lambda CDM}(z)]$ following Equation \ref{eq:ISW_definition2}.

Furthermore, we also tested the consistency of our ISW-based results with relevant other constraints on the growth rate of structure at the redshift range probed by the eBOSS survey. In the case of the FastSound results at $z\approx1.4$ \citep[][]{Okumura2016}, eBOSS DR16 consensus results from clustering analyses \citep{Alam2021}, and constraints from the eBOSS QSO voids \citep[][]{Aubert2020}, we \emph{converted} the measured values of the growth parameter combination $f\sigma_{8}(z)$ to a constraint on $f(z)$ by calculating $f^{*}(z)=f\sigma_{8}(z)/\sigma_{8}^{Pl}(z)$ with the assumption of a \emph{Planck} 2018 cosmology. We stress that this simplistic comparison and scaling are \emph{not} sufficient to draw conclusions. Nonetheless, they are useful to examine possible common trends in different observations of the growth rate of structure.

As shown in Figure \ref{fig:figure_7}, the eBOSS emission line galaxies (ELG) show good agreement with our results from our lowest redshift bin at $0.8<z<1.2$, although the $z$ range overlap is only partial between them. At $1.2<z<1.5$, the growth rate constraint from the FastSound survey is again perfectly consistent with our ISW-based estimation of $f$. 

Most relevantly, we also see good agreement with the consensus results from the eBOSS QSO growth rate analyses \citep{Hou2021,Neveux2020} and the less constraining but methodologically more similar result from cosmic voids in the eBOSS QSO survey at $z\approx1.5$ \citep{Aubert2020}. Here we note that these results are based on a single-bin analysis of the eBOSS QSO catalogue, and therefore they are not sensitive to changes in $f(z)$ that we observed at $z\approx1.5$ compared to the expected $\Lambda$CDM evolution. 

At even higher redshifts ($z\approx3$), the $f=1.46\pm0.29$ constraint from the SDSS Lyman-$\alpha$ forest also appears to be consistent with a stronger gravitational growth at ``cosmic noon'' \citep{McDonald2005}, which may provide further insight on this problem if measured with higher precision.

We conclude that, despite the criticism by \cite{Hang20212pt}, the AvERA \emph{toy-model} approach might provide valuable insights at least in the context of the evolution of supervoids; even though it might not be the final answer for the ISW anomalies in general. Certainly, up-coming data from the Euclid \citep[][]{euclid}, DESI \citep[][]{DESI}, and J-PAS \citep[][]{jpas2014}, surveys will provide tighter constraints on these anomalous ISW signals from supervoids. Furthermore, complementary analyses using superclusters, and more detailed measurements of the CMB lensing imprint of super-structures \citep[see e.g.][]{Vielzeuf2019,Raghunathan2019} at various redshifts may uncover additional details about the ISW anomalies, and their possible relation to other interesting puzzles in cosmology \citep[see e.g.][]{Riess2019,Heymans2021}.

\section*{Acknowledgments}

The authors thank Marie Aubert, Julian Moore, Carlos Hern\'andez-Monteagudo for their insightful comments and suggestions which improved the clarity of the manuscript.

AK has been supported by a Juan de la Cierva fellowship from MINECO with project number IJC2018-037730-I, and funding for this project was also  available in part through SEV-2015-0548 and AYA2017-89891-P. IC and GR acknowledge support from National Research, Development and Innovation Office of Hungary through grant OTKA NN 129148. IS acknowledges support from the National Science Foundation (NSF) award 1616974.

\section*{Data availability}

The eBOSS QSO data\footnote{https://www.sdss.org/dr16/}, the MXXL halo mock catalogues\footnote{https://tao.asvo.org.au/tao/}, and the CMB temperature\footnote{https://www.cosmos.esa.int/web/planck} and lensing\footnote{http://www.cosmostat.org/products} maps are publicly available. The simulated ISW analysis software\footnote{https://github.com/beckrob/AvERA\_ISW} and the AvERA simulation tools\footnote{https://github.com/eltevo/avera} which provided important foundations for this article are also available from public websites, or will be shared on reasonable request to the corresponding author.

\bibliographystyle{mnras}
\bibliography{refs}
\end{document}